\documentclass[twocolumn,amsmath,amssymb,showpacs,superscriptaddress,prb,aps,longbibliography,10pt]{revtex4-1}
\usepackage{graphicx}
\usepackage{bm}
\usepackage{dcolumn}
\usepackage{amssymb}
\usepackage{amsmath}
\usepackage{amsfonts}
\usepackage{epsfig}
\usepackage{graphicx}
\usepackage{stackrel}
\usepackage{paralist}
\usepackage[rgb]{xcolor}
\usepackage{todonotes}
\usepackage{pdfcomment}
\usepackage{mathrsfs}
\newcommand{\I}{\mathrm{i}}
\newcommand{\refEq}[1]{Eq.~(\ref{#1})}
\newcommand{\refFig}[1]{Fig.~\ref{#1}}

\newcommand{\citeRef}[1]{Ref.~\onlinecite{#1}}
\newcommand{\Conv}{\mathop{\scalebox{1.5}{\raisebox{-0.2ex}{$\ast$}}}}%
\DeclareMathOperator{\sinc}{sinc}
\newcommand{\bl}[1]{\mathbf{#1}}
\newcommand{\mean}[1]{\langle{#1}\rangle}
\def\be{\begin{equation}}
\def\ee{\end{equation}}
\def\ba{\begin{align}}
\def\ea{\end{align}}

\newcommand{\bmat}{\begin{bmatrix}}
\newcommand{\emat}{\end{bmatrix}}
\newenvironment{psmallmatrix}
  {\left(\begin{smallmatrix}}
  {\end{smallmatrix}\right)}
	
\begin{document}
\title{Current at a distance and resonant transparency in Weyl semimetals}
\author{Yuval Baum}
\affiliation{Department of Condensed Matter Physics, Weizmann Institute of Science, Rehovot 76100, Israel}
\author{Erez Berg}
\affiliation{Department of Condensed Matter Physics, Weizmann Institute of Science, Rehovot 76100, Israel}
\author{S. A. Parameswaran}
\affiliation{Department of Physics and Astronomy, University of California, Irvine, CA 92697, USA}
\author{Ady Stern}
\affiliation{Department of Condensed Matter Physics, Weizmann Institute of Science, Rehovot 76100, Israel}
\begin{abstract}
Surface Fermi arcs are the most prominent manifestation of the topological nature of Weyl semimetals. In the presence of a static magnetic field oriented perpendicular to the sample surface, their existence leads to unique inter-surface cyclotron orbits.
We propose two experiments which directly probe the Fermi arcs: a magnetic field dependent non-local DC voltage and sharp resonances in the transmission of electromagnetic waves at frequencies controlled by the field. We show that these experiments do not rely on quantum mechanical phase coherence, which renders them far more robust and experimentally accessible than quantum effects.  We also comment on the applicability of these ideas to Dirac semimetals.
\end{abstract}
\maketitle

\section{Introduction}
Topology in various guises plays a central role in modern condensed matter physics~\cite{Top1,Top2}. In recent years, a sharpened understanding of the topology of electronic wave functions in crystals has stimulated the discovery of new phases of matter~\cite{Kitaev,class}.
Among the remarkable manifestations of these phases are robust gapless edge modes and precisely quantized bulk response functions, linked by the celebrated bulk-boundary correspondence.
Although the original applications of topological ideas to band structures relied on the existence of a fully gapped bulk spectrum, more recently it has been recognized that protected surface states can arise even in gapless systems~\cite{gapless1,gapless2,gapless3,WeylRevTurner,WeylArcsWen,WeylFromTI,Hosur}.

The prototypical example of a gapless topological phase is a Weyl semimetal (SM)~\cite{WeylRevTurner,WeylArcsWen,WeylFromTI,Hosur}:
a three dimensional crystalline material where the bulk is gapped except at an even number of points in the Brillouin zone (BZ) in which the energy bands touch --- the  Weyl nodes.
In the vicinity of these nodes, the electrons disperse as massless relativistic particles, and may be described at low energies by a Weyl Hamiltonian familiar from particle physics, $\mathcal{H}\approx \pm \hbar v_{\rm 0}{\mathbf k}\cdot {\mathbf \sigma}$. Here ${\mathbf \sigma}$ is a pseudo-spin degree of freedom, $\bf k$ is measured from the Weyl node, and the sign is set by the electron chirality~\footnote{The general form of the Weyl Hamiltonian is $k_iA_{ij}\sigma_j+f(\textbf{k})\sigma_0$, where $A$ is a $3\times3$ matrix and $f$ is a linear function of $k$. Our results should not depend on the exact form of $A$ and $f$, at least qualitatively. For simplicity, we consider only the isotropic case.}.
From the perspective of band topology, a Weyl node is either a source or a sink of Berry flux, depending on the chirality. While the total Berry flux in the $3$D BZ must be zero, as long as Weyl nodes of opposite chirality are separated in momentum space, there must exist two-dimensional cuts of the 3D BZ in which the Chern number is non-zero. Each such cut defines a 2D integer quantum Hall state. For a finite sample, these states necessarily have edge modes on the appropriate surfaces. Consequently, for 3D samples these real-space surfaces host ``Fermi arcs" of states that belong to the two dimensional momentum-space Fermi-surface~\cite{WeylArcsWen}.

When a magnetic field is applied perpendicular to real-space surfaces that carry Fermi arcs, electrons traverse unique cyclotron orbits that connect opposite surfaces of the sample. These cyclotron orbits are central to our discussion. They are reflected in the quantum Shubnikov-deHaas (SdH) oscillations of the resistance~\cite{QuantumOsc}, as was recently observed in a closely related Dirac semimetal Cd$_3$As$_2$ \cite{QuantumOscMeasur}. In addition, Weyl semimetals have been predicted to exhibit various unusual magneto-transport phenomena, related to the ``chiral anomaly''~\cite{Son2013, WeylNonLocalTransport}.

Following an early suggestion~\cite{WeylPyroc} that certain iridium pyrochlores may host a semimetal with $N_W=24$ Weyl nodes, the number of Weyl SM candidates has proliferated to include HgCr$_{\rm 2}$Se$_{\rm 4}$ with $N_W=4$~\cite{WeylMatHg} and heterostructures of normal and magnetically doped topological insulators~\cite{WeylFromTI}. In particular, non-centrosymmetric transition-metals, such as TaAs, have been predicted in~\citeRef{tranmetal1,tranmetal2} to be Weyl SMs. Indeed, recent photoemission and transport measurements provide strong evidence for realization of a Weyl SM phase in TaAs~\cite{TaAs,TaAs2,TaAs3,TaAs4,TaAs5}. In addition, closely related Dirac semimetals, that also carry Fermi arcs, have been observed experimentally~\cite{WeylExp1,WeylExp2,WeylExp3,WeylExp4,WeylExp5}.

In this work we show how inter-surface cyclotron orbits affect the electronic properties of Weyl semimetals already at the semi-classical level. As a result, we are able to propose two experiments to probe these trajectories without requiring quantum mechanical phase coherence. These experiments pose far less stringent requirements than SdH oscillations in terms of sample  purity, surface roughness and temperature.

\begin{figure}
\centering
\includegraphics[width = \linewidth]{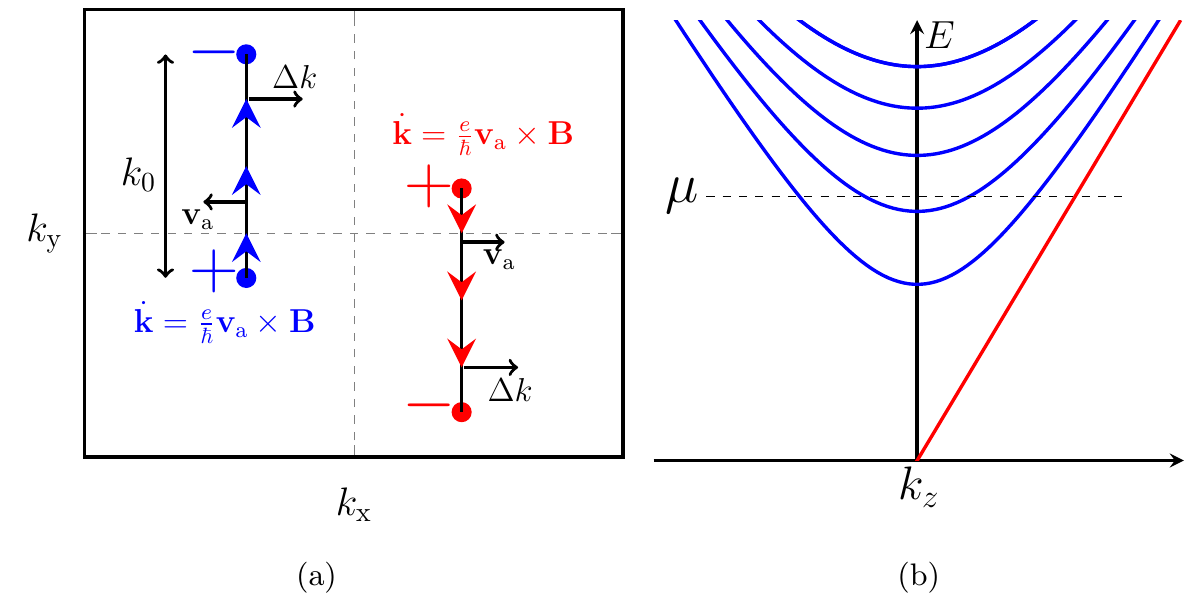}
\caption{(a) Fermi arcs in the surface Brillouin zone for a time-reversal-invariant Weyl SM. For simplicity, we ignore the curvature of the arcs and consider straight arcs that are directed along the $y$-axis. The extent of the arcs is $k_0$. The red dots denote pair of Weyl nodes with a positive/negative chirality. The blue dots denote their time-reversal partners. The electron velocity, $\mathbf v_{\mathbf a}$, is perpendicular to the arc at each point. Within the straight arcs approximation, the magnitude of $|\mathbf v_{\mathbf a}|\equiv v_a$ is independent of $\textbf{k}$. In the presence of a perpendicular magnetic field, the surface electrons 'slide' along the Fermi arcs towards the negative chirality Weyl node. An electric field pulse in the x-direction leads to a momentum shift, $\Delta k$, i.e., the right arc is slightly populated while the left arc is depopulated. (b) Bulk LL near one of the Weyl nodes. The red line denotes the chiral LL.\label{arcs}}
\end{figure}

We consider a box-shaped slab of Weyl semimetal with the main axes being the Cartesian axes, and with the Fermi arcs on the $z=0,L$ surfaces. We assume $L$ to be much smaller than the other two dimensions. A magnetic field \textbf{B} is to be applied in the $z$-direction. For concreteness, we consider a time-reversal symmetric Weyl SM. Therefore, the surface must include an even number of Fermi arcs, and the minimal number of Weyl nodes is $N_{\rm W}=4$, as shown in~\refFig{arcs}a.
For simplicity, we ignore the curvature of the arcs and consider straight Fermi lines directed in the $y$ direction with a constant band velocity $v_a$ along the $x$ direction and momentum extent $k_0$. Although the electrons that reside on a single arc posses a non-zero mean velocity, the surface current is zero, due to the cancellation between any pair of time reversed arcs.

We propose two related experiments. In the first, we consider two parallel line-shaped Ohmic contacts placed on the $z=L$ surface of the slab, separated by $a\ll L$ as depicted in~\refFig{Setup}a. In standard metals, the current path extends a distance of the order of $a$ into the bulk. Hence, for $a\ll L$ the current at the $z=0$ surface vanishes as $a/L^2$. We show that in the presence of a perpendicular magnetic field when a voltage $V$ is applied between the two contacts on the top surface, opposite currents are induced in the two surfaces of the sample. As a consequence, a voltage $\alpha V$ is induced in the bottom surface, that scales linearly, $\alpha\propto |\textbf{B}|$, for small fields.

Our second proposal considers an electromagnetic microwave radiation propagating from $z=\infty$ downward along the $-z$-axis, as depicted in~\refFig{Setup}b. We show that when the slab is much thicker than the semimetal skin depth, such that the radiation is expected to be mostly reflected by the slab, there are transmission resonances at which a significant part of the radiation is transmitted, with an amplitude that is independent of $L$. The transmission amplitude is again linear in $|\textbf{B}|$ for small fields, as is the resonant frequency.

The effects we discuss here involve transfer of electrons between Weyl nodes. In a clean Weyl semi-metal in a magnetic field, there are two mechanisms for electrons to be transferred between different nodes: the chiral anomaly (which is effective in the bulk) and the Fermi arcs at the surface. Our proposals rely on the second mechanism, in contrast to those of \citeRef{WeylNonLocalTransport} that originate from the first. 

\section{Non-local conductivity}
Both these phenomena originate from the same mechanism
--- the non-local conductivity of a Weyl semimetal in a magnetic field. Linear response theory defines the conductivity in space-time through the relation $j({\bf r},t)=\int d{\bf r'}dt'{\tilde\Sigma}({\bf r},{\bf r'}, t-t'){\bf E}({\bf r'},t')$. Applying Drude theory to a doped Weyl/Dirac node, we find
\begin{equation}
{\tilde\Sigma}\equiv {\tilde \Sigma}_D=\frac{e^2}{h}v_0k_F^2 e^{-t/\tau}\delta({\bf r}-{\bf r'})\equiv\tilde{\sigma}_0\delta({\bf r}-{\bf r'}),
 \label{Drudecond}
 \end{equation}
where $\tau$ is the momentum relaxation time, $v_0$ the velocity, $k_F$ the Fermi momentum and $\tilde{\sigma}_0$ is the local Drude conductivity. We use $\tilde\Sigma$ to denote time-domain conductivity, and remove the tilde for frequency-domain conductivity. 

\begin{figure} [t]
\centering
\includegraphics[width = \linewidth]{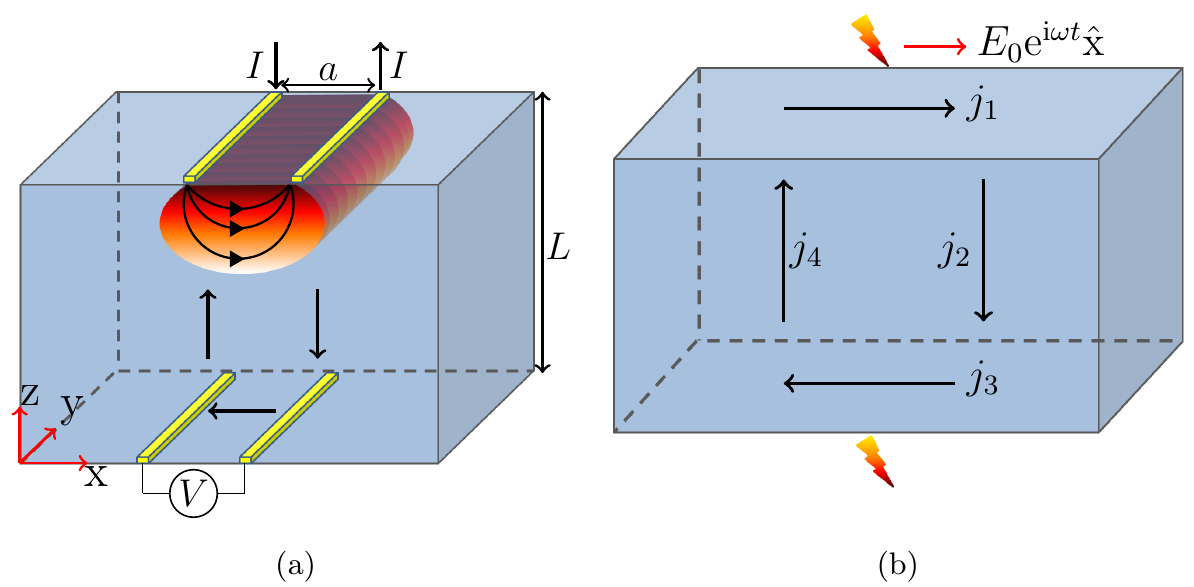}
\caption{(a) DC configuration: a current, $I$, is injected via the upper left contact and withdrawn from upper right contact. The voltage difference between the lower contacts is measured. Without the non-local orbits, current flow in a 'cigar' shape configuration between the contact. (b) AC configuration: current cycle due to a pulse of EM field on the top surface leads to emission from the lower surface. \label{Setup}}
\end{figure}

The unique inter-surface cyclotron orbits result in a non-local contribution to the conductivity $\tilde\Sigma$, not captured by Eq.\eqref{Drudecond}. Consider an electric field pulse induced by a vector potential $A_x(z,t)=-A_0f(z)\theta(t)$, where $f(z)$ is spread over a short length scale $d$ around $z=L$. This length scale is a few times the spatial extent of the surface states in the $z$-direction. Immediately after the pulse, a surface current emerges, as may be computed from the shift of the Fermi level of the two Fermi arcs due to the applied vector potential.
The induced current density is $j_0=e v_a k_0\Delta k$ where $\hbar\Delta k=eA_0f(z)$.

Following this shift of the Fermi surface, the magnetic field $\mathbf{B}$ causes the surface electrons to `slide' along the Fermi arcs towards the negative chirality Weyl point, at a rate $\dot{\mathbf k}=\frac{e}{\hbar}\mathbf v_{\mathbf a}\times\mathbf B$ (\refFig{arcs}a). As electrons slide on the Fermi arcs they eventually arrive at the Weyl nodes, where they merge into the $3$D bulk. In the presence of the magnetic field, the bulk spectrum in the vicinity of each Weyl node consists of dispersive Landau levels (LL) as depicted in \refFig{arcs}b. Specifically, the chiral LL that originates from the $\nu=0$ LL of Dirac electrons serves as a direct passageway for electrons from the top to the bottom surface.
Note that the electrons travel along the same cyclotron orbit responsible for the unique quantum oscillation signature of Weyl SMs~\cite{QuantumOsc}.

When the excess current is all in the chiral state, i.e., when the Fermi energy $\mu$ satisfies $|\mu|<v_0\sqrt{2\hbar eB}$, the entire current flows to the opposite surface, where it flows in the $-{\hat x}$ direction. In the $z=0$ layer the sign of the velocity reverses, and hence so does the sign of ${\dot k}_y$, such that the electron motion along the arc brings it to a chiral state that flows back to $z=L$, completing the inter-layer cyclotron motion. In the absence of scattering, this cycle repeats indefinitely. Scattering between different Weyl nodes is detrimental to inter-surface cyclotron orbits. In clean samples such scattering is rare due to the large momentum difference between nodes. Scattering within a Weyl node is highly suppressed when $|\mu|<v_0\sqrt{2\hbar eB}$ since there are no Landau levels to scatter into; we discuss the case $|\mu|>v_0\sqrt{2\hbar eB}$ below. 

Altogether, then, a pulse of an electric field in the $x$-direction at $z=L$ leads to alternating and opposite currents $j_1,j_3$ in the two surfaces, and an alternating bulk current $j_2-j_4$ between the surfaces. The periodic orbit is depicted in \refFig{Setup}b. The period $T_{\rm 0}$ is obtained by combining two basic timescales: $T_{\rm 0}=2(T_{\rm z}+T_{\rm arc})$, where
$T_{\rm z}=L/v_{\rm 0}$ is the time needed for electrons to cross from the upper to the lower surface, and $T_{\rm arc}=\hbar k_{\rm 0}/(eBv_{\rm a})$ is the time needed for electrons to slide along the Fermi arc. Here again, $k_0$ is the extent of the arc in $k$-space and $v_a$ is the magnitude of the arc velocity as depicted in \refFig{arcs}a.

We now focus on the current in the lower surface, $j_3$ in \refFig{Setup}b, from which we calculate ${\tilde\Sigma}(0,L,t-t')$. We consider only the evolution of the excess electrons due to the imbalance generated by the electromagnetic (EM) pulse, since the current is solely determined by these electrons. At $t=0$ the excess electrons populate all the states along the Fermi arc. These states are gradually depleted to the bulk at a constant rate, $\frac{e}{\hbar}Bv_{\rm a}$, until all states are empty at $t=T_{\rm arc}$. The first excess electron reaches the lower surface at $t=T_{\rm z}$. Gradually, more and more excess electrons reach the lower surface until $t=T_{\rm z}+T_{\rm arc}$, where a maximum in the current density, $j_{\rm 3}$, occurs. Then, the excess electrons start to leave the lower surface and move towards the upper one, completing the cycle. 
Scattering between the Weyl nodes is expected to suppress this current, and we characterize it by
a relaxation time $\tau_v$ and length $l_v$. The current decays as $e^{-t/\tau_v}$ as more and more cycles occur. The effect of intra-node scattering, which takes place when $|\mu|>v_0\sqrt{2\hbar eB}$ is discussed below.

\begin{figure}
\centering
\includegraphics[width = \linewidth]{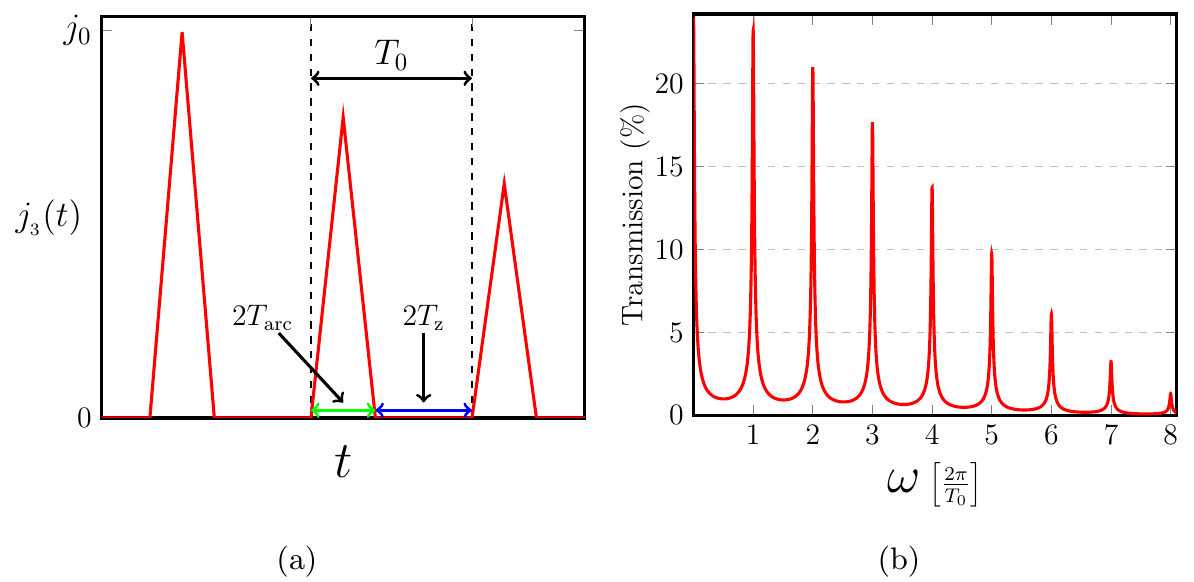}
\caption{(a) A sketch of real time current on the lower surface for $|\mu|<v_0\sqrt{2\hbar eB}$. In this case, the source of current decay is inter-valley scattering. (b) The transmission coefficient as a function of the source frequency. A significant part of the radiation is transmitted whenever the frequency of the applied EM field is an integer multiple of $\omega_0=2\pi/T_0$. \label{fig1}}
\end{figure}

A sketch of the real-time current on the lower surface, $j_3(t)$, is shown in \refFig{fig1}a for the case $|\mu|<v_0\sqrt{2\hbar eB}$. Assuming a linearly rising current as the electrons arrive at the bottom surface, the current on the lower surface is obtained by convolving a periodic function representing the cycles, with a `triangle function' $\Lambda(t)$ describing the growth and decay of the current in each cycle, and scattering induced exponential suppression.
\begin{equation}\label{bot_curr}
j_3(t)\approx \Lambda(t)\Conv \sum\limits_{n=0}^{\infty}\delta\left(t-T_0\left(n+\frac{1}{2}\right)\right)e^{-\frac{t}{\tau_v}}
\end{equation}
where $\Conv$ denotes a convolution. The current induced at the same surface at which the electric field is applied is given by a similar expression, with the $1/2$ absent.

Fourier transforming and substituting $j_0$ we get an expression for contribution of the cyclotron orbits to the conductance,
\begin{align}
&\Sigma(0,L,\omega)\equiv\frac{-j_x(0,\omega)}{E_x(L,\omega)}=\frac{\I G_0\sinc^2\Big(\frac{\omega T_{\rm arc}}{2}\Big)}{2\sinh{(\varphi)}},
\\
&\Sigma(L,L,\omega)\equiv\frac{j_x(L,\omega)}{E_x(L,\omega)}=\Sigma(L,0,\omega)e^{\varphi},
\label{nlconduc}
\end{align}
where $\varphi=(\tau_v^{-1}-\I\omega)T_0/2$ and  $G_0=e^2v_{\rm a}k_0T_{\rm arc}/h d$.
Consequently, we approximate the non-local currents by
\begin{equation} \label{nl}
j_x^{\rm NL}\approx d\sigma_{W}(\omega)\delta(z-L)\left[E_x(L)e^{\varphi}-E_x(0)\right]+(0\leftrightarrow L),
\end{equation}
where $\sigma_{W}\equiv\Sigma(0,L,\omega)$ is the `Weyl conductivity'. The Drude conductivity $\sigma_0$ and the Weyl conductivity $\sigma_{W}$ have the same units.
In our estimate of $\sigma_W$ we neglect the variation along the arc of the localization length of surface states in the $z$-direction. 
 We also ignored diabatic transitions into the bulk \cite{QuantumOsc}. The latter effect is just a correction $k_0\rightarrow k_0 - \beta \ell_B^{-1}$ where $\ell_B$ is the magnetic length and $\beta \sim \mathcal{O}(1)$. 

\section{Experimental Signatures of Non-local Conductivity}

Having calculated the non-local part of the conductivity, we are in a position to analyze the two experiments we propose.

\subsection{DC Transport}
\begin{figure}[h]
\centering
{\includegraphics[width = \linewidth]{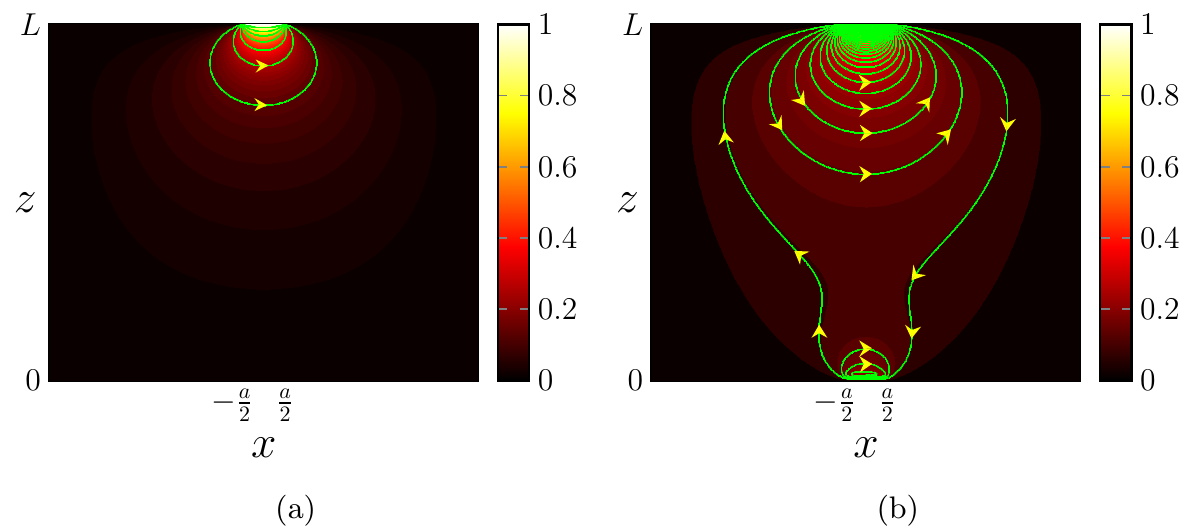}}
\caption{The stream function $\psi$ for (a) $\rho_W/\rho_0=0$ and (b) $\rho_W/\rho_0=\sigma_W/\sigma_0=0.15$. The current flows along contours (green) of constant $\psi$ (in the direction of the arrows), and its magnitude is given by $|\nabla\psi|$ which is proportional to the density of the green contours. In (a) we find that $ \frac{V_{CD}}{V_{AB}}\approx 0.005$ while in (b) we find $\frac{V_{CD}}{V_{AB}}\approx -0.273$.\label{fig_app2}}
\end{figure}

When a DC voltage is applied between the lines $x=\pm a/2$ on the $z=L$ surface, a current flows in the sample. The current density and electric field must satisfy Kirchoff's rules and Ohm's law, 
\begin{align}
 \label{kirch}
&\bl{\nabla}\cdot\bl{j}=0 \\
&\bl{\nabla}\times\bl{E}=0 \\
&\bl{j}=\Sigma \bl{E}
\end{align}
where $\Sigma$ is the calculated conductivity (including the non-local part in (\ref{nlconduc})). We ignore the inter-valley currents due to the chiral anomaly, that would generically make the bulk conductivity moderately anisotropic (see the next section for a discussion). 
The boundary conditions impose zero current perpendicular to the surface everywhere except at the contacts, and enforce the voltage $V=-\int {\bf E}\cdot{\bf dl}$ between the contacts. At zero magnetic field the conductivity is purely local. Consequently, the current path extends a distance of the order of $a$ into the bulk, and vanishes as $a/L^2$ at the $z=0$ surface. Fig.~\ref{fig_app2} presents the current path in the presence of a non-local conductance. The color represents the stream function $\psi(x,z)$ which is related to the current as follows:
\be
J_x=-\partial_z\psi,\, J_z=\partial_x\psi.
\ee
A few equal value contours of $\psi(x,z)$ and the direction of the current along them are also shown in to Fig.~\ref{fig_app2}.
The current flows along contours (green) of constant $\psi$ (in the direction of the arrows). The values of the presented contours are equally spaced. Hence, the current that flows between any two contours is the same, and the magnitude of the current density is proportional to the density of the (green) contours.
As can be seen in Fig.~\ref{fig_app2}a, only a small current reaches the lower surface. On the other hand, as seen in Fig.~\ref{fig_app2}b, the application of a magnetic field leads to an opposite current at the $z=0$ surface. 
The chiral modes in the bulk, which mediate the current between the two surfaces, are always parallel to the magnetic field. Therefore, when the magnetic field is along the z-axis, the current on the lower surface appears approximately below regions on the upper surface where a non-negligible electric field is developed.   
Details of the calculation are given in Appendix~\ref{app1a}.

\subsubsection{Inter-valley currents due to chiral anomaly} 
We address the effect of the chiral anomaly on the DC effect. When a DC voltage is applied between the two upper contacts, the electric field lines bend into the bulk at the vicinity of the upper surface. The resulting $E_z$ leads to a density transfer between the two chiral modes in the bulk according to:
\be
\frac{dn}{dt}=\left(\frac{e}{h}\right)^2\bl{E}\cdot \bl{B}=\left(\frac{e}{h}\right)^2E_zB
\ee
This is the so-called chiral anomaly. The change in the bulk current is then given by: $\frac{dj_z}{dt}=ev_0\frac{dn}{dt}$. Employing theses relations, we conclude that an extra current $j_z$ is produced due to the chiral anomaly:
\be  \nonumber
j_z^{\rm anom}\sim\frac{dj_z}{dt} T_z=\frac{dj_z}{dt}\frac{L}{v_0}=\left(\frac{e}{h}\right)^2ev_0B\frac{L}{v_0}E_z=\frac{e^2}{h}\frac{L}{\ell_B^2}E_z
\ee
where $\ell_B$ is the magnetic length. Evidently, the chiral anomaly renormalizes the longitudinal conductivity in the $z$ direction, i.e., $\sigma_{zz}=\sigma_0+\frac{e^2}{h}\frac{L}{\ell_B^2}$, while the other directions remain unchanged, $\sigma_{xx}=\sigma_0=\frac{e^2}{h}k_F^2\ell$. Hence, the effect of the chiral anomaly is to make the bulk conductivity anisotropic.
The solution of the Kirchoff-Ohm's equations in the anisotropic case may be related to the solution in the isotropic case by the following simple rescaling:
\begin{align} \nonumber
&(E_x,E_z)=(\tilde{E_x},\alpha^{-1}\tilde{E_z}),\\ \nonumber
&(j_x,j_z)=(\tilde{j_x},\alpha\tilde{j_z}),\\ \nonumber
&(x,z)=(\tilde{x},\alpha\tilde{z}),\,\mbox{and }\sigma_W=\alpha\tilde{\sigma}_W \nonumber
\end{align}
where the tilde represent quantities evaluated in the isotropic case, and $\alpha=(\sigma_{zz}/\sigma_{xx})^{1/2}\approx((2N+1)^{-1}L/\ell)^{1/2}$ with $N$ being the filling factor of the non-chiral Landau levels. Thus, as long as $\alpha\sim\mathcal{O}(1)$ the DC effect is only moderately affected by the chiral anomaly.

\subsection{AC Transport}
The second experiment examines the transmission of electro-magnetic radiation through the slab. Assuming translational invariance in the $xy$ plane, the propagation of a monochromatic electromagnetic wave along the $z$-direction is described by the inhomogeneous wave equation ($c=1$),
\begin{equation} \label{Max2}
\left(\partial_{z}^2+\omega^2\right)\mathbf{E}=-\left(\I\omega\mu \mathbf{j}-(\partial_z\rho)\hat{z}\right)
\end{equation}
where $\rho$ and $\mathbf{j}$ are the charge and current densities inside the metal and $\mu$ is the permeability of the metal. Assuming incident radiation polarized along $x$ and with an amplitude $E_0$, only the $x$-component of the electric field is nonzero outside the slab, so that the solution of \refEq{Max2} satisfies $\mathbf{E}(\omega,z=0)=E(0)\hat{\mathbf{x}}$ and $\mathbf{E}(\omega,z=L)=E(L)\hat{\mathbf{x}}$.
Solving for $E(0)$ as a function of $E_0$ and $\omega$ allows us to characterize the transmission amplitude of the wave through the sample, $|E(0,\omega)/E_0|$.

In the presence of a non-local electrodynamic response, we can rewrite
\refEq{Max2} as
\begin{equation} \label{Max3}
\left(\partial_{z}^2+\omega^2+\I\omega\mu\sigma_0\right)E_x(z,\omega)=\I\omega\mu j^{\rm NL}
\end{equation}
where $\sigma_0$ is the conductivity of the metal, and $j^{\rm NL}$ is defined in \refEq{nl}. For $\Sigma=0$, the solution of \refEq{Max3} is an evanescent wave decaying exponentially into the bulk, $E_x(z)\sim e^{-z/\delta}$ with the skin depth $\delta^{-1}=\mbox{Im}(\omega^2+\I\omega\mu\sigma_0)^{1/2}\approx(\omega\mu\sigma_0/2)^{1/2}$. We assume that the bulk is sufficiently metallic that the Hall angle is small; then, in the absence of non-local terms, the transmission is exponentially small in the sample width.

Solving \refEq{Max3}, we find a significant increase in the transmission coefficient when the frequency of the applied EM field is an integer multiple of $\omega_0=2\pi/T_0$, as depicted in \refFig{fig1}b.
In appendix~\ref{app2}, we analyze the transmission coefficient as a function of the ratio $\sigma_{W}/\sigma_{0}$.
We find that for frequencies where $\sigma_{W}(\omega)$ is non-negligible, electric field $E(0)$ is comparable to $E(L)$.
As $\sigma_{W}$ increases, the transmission increases rapidly until $\sigma_{W}\sim \sigma_0$, where it saturates. The saturation value depends on the ratio $\omega/\sigma_0$.
The non-local orbits ``connect the surfaces'', which eventually leads to a field in the lower surface $E(0)$ that may be as high as the field $E(L)$. That does not amount to a full transmission because $E(L)$ is not $E_0$.
For a thick standard metal the radiation is partially reflected and partially absorbed. The absorption originates from the penetration of an electric field into the metal.
In the Weyl case, the mechanism we described transforms part of the absorption into transmission, with an effectiveness that increases with $\sigma_W/\sigma_0$. However, this mechanism does not completely eliminate the reflection, and hence the transmission saturates at a value smaller than one. The increase in absorption with $\omega$ in standard metals is converted into resonant transmission at multiples of the cyclotron frequency.

In other words, in the presence of a magnetic field the Weyl semimetal acts as a ``narrow band-pass filter'' for incident EM radiation around harmonics of the frequency $\omega_{\rm 0}=2\pi/T_{\rm 0}$. The frequency $\omega_{\rm 0}$ can be tuned easily, since it depends on both the system width, $L$, and the magnetic field $B$. In crystals where the surface hosting Fermi arcs lacks $C_4$ symmetry, the results of the AC experiment would be polarization dependent.
More details appear in appendix~\ref{app2}.

The resonant transmission we find resembles the Azbel'-Kaner cyclotron resonance \cite{AzbelKaner1,AzbelKaner2,AzbelKaner3,AzbelKaner4,AzbelKaner5}, in which magnetic fields {\it parallel} to the sample surface lead to enhanced transmission when the cyclotron radius and frequency match the sample thickness and the frequency of the applied electromagnetic field, respectively. Note that previously proposed nonlocal effects in Weyl SMs~\cite{WeylNonLocalTransport} are quite distinct from those considered here.

\section{Effects of scattering}
Shubnikov-deHaas oscillations of the resistance are a consequence of quantum interference, and are hence suppressed by elastic and inelastic scattering, even at small momentum transfer. As such, they are suppressed exponentially in the ratio of the size of the cyclotron orbit to the mean free path and the  ratio of the temperature to the cyclotron energy \cite{Lifshitz}. 
 For the inter-surface cyclotron orbits we consider this requirement restricts the sample size $L \lesssim l$, achieved through low temperature measurements on very thin, high-purity samples.

In contrast, the two transport phenomena that we discuss here, while they rely on the existence of the cyclotron orbits, do not rely on quantum interference around the orbits, and therefore phase coherence is unimportant. Their existence is affected instead by how multiple scattering events in the bulk alter the structure of $\tilde{\Sigma}(t)$ (shown in \refFig{fig1}a).
To address this, we return to the spectrum depicted in \refFig{arcs}b, and assume the chemical potential $\mu$ to be such that the chiral state overlaps in energy with $N>1$ Landau levels.

For concreteness, we consider a Weyl node where the chiral state flows downwards (in the $-z$ direction). In the absence of scattering the Weyl node has $N+1$ modes flowing downwards and $N$ modes flowing upwards. We model the semiclassical scattering as a diffusion process, in which the electron is scattered between the $2N+1$ modes, with a scattering event taking place at a rate $1/\tau$, with $\tau\ll T_z$. For each Landau level, the upwards and downwards velocities are identical in magnitude, but velocities are not necessarily identical between Landau levels. Due to the unpaired chiral state, the diffusion process results in a downwards average drift velocity $v_d$, which is smaller than the velocity of the chiral mode, $v_0$. Neglecting velocity differences between different Landau levels, $v_d=v_0/(2N+1)$, and the average crossing time between surfaces becomes $\langle T_z\rangle=\frac{L}{v_d}> \frac{L}{v_0}$. Moreover, the diffusion  leads to fluctuations in the crossing time, characterized by a variance $\Delta T_{\rm z}$ around $\mean{T_{\rm z}}$. The relative width of the distribution of arrival times to the bottom surface is given by
\begin{equation}
\frac{\Delta T_{\rm z}}{\mean{T_{\rm z}}}=\frac{\sqrt{D\mean{T_{\rm z}}}}{v_d \mean{T_{\rm z}}}=\sqrt{\frac{(2N+1)l}{L}},
\end{equation}
where $D=v_0^2\tau$ is the diffusion constant, and $l=v_0\tau$ is the mean free path. Thus, as long as $l(2N+1)\ll L$, the first passage of the current from one surface to another occurs at a rather well defined time. In terms of ${\tilde\Sigma}(t)$, this will lead to a shift, smoothening and broadening of the first peak of \refFig{fig1}a. The area under the peak stays constant, since the entire current crosses from one surface to another (see below).
As the current continues along the inter-surface cyclotron orbits the spread of the crossing times increases, and hence the peaks of ${\tilde\Sigma}(t)$ further broaden, until effectively merging together at a time $t$ where $Dt\approx L^2$.

Interestingly, intra-Weyl node scattering does not suppress the cyclotron current that gives rise to the non-local $\tilde\Sigma(t)$, even after the peaks merge together. The reason for that may be understood by considering an analog situation in the realm of the two dimensional quantum Hall effect. Imagine a $\nu=1$ QHE  state at the half-plane $x<0$, with an downwards-moving chiral edge along the $y$-axis. Now imagine coupling the $y<0$ part of the edge to $N$ semi-infinite quantum wires of spinless electrons, each carrying one upwards-moving chiral mode and one downwards-moving chiral mode. Independent of whether the coupling is ordered or random, it cannot block a downwards moving current emanating from $y=+\infty$, due to the chirality of the QHE edge.
The coupling to the wires merely renormalizes the velocity of the edge state and modifies its wave function.

In Weyl semimetals the surface plays the role of the QHE $y>0$ edge. As long as the surface is free of scattering, the momentum of states on the surface is well defined. The finiteness of the arc, together with the drift imposed by the equation of motion $\hbar\dot{\mathbf k}=e\mathbf v_{a}\times\mathbf B$, enforce a flow of the electrons into the bulk, which plays the role of the $y<0$ region in the QHE case. This flow cannot be reversed as long as electrons are not transferred to a Weyl node with an opposite chirality, since the direction of $\dot{\mathbf k}$ is fixed for every arc. Thus, the entire current that enters the bulk from one surface must cross all the way to the other surface. As long as electrons' chirality changes only at the surface, the peaks in \refFig{fig1}a preserve their area under scattering.

Transforming these observations into the frequency domain, we find that $\Sigma(\omega)$ shows clear resonances, as in \refFig{fig1}b, around harmonics of a modified cyclotron frequency $2\pi/\mean{T_0}$. The height of the peaks decays fast with frequency, but the low frequency peaks remain almost unaltered when compared to the clean case. The DC part, $\Sigma(\omega=0)$, which involves the integral over all times, is not altered  by intra-node scattering. Consequently, intra-node scattering does not affect the signal in the first experiment we proposed here. Two factors that do affect this experiment are inter-node scattering and the chiral anomaly. Inter-node scattering is characterized by valley relaxation length, $l_v$, which can be tens of microns~\cite{WeylNonLocalTransport}. It should therefore be rather ineffective.
For the AC experiment, inter-valley scattering suppresses the area under the peaks in ${\tilde\Sigma}(t)$ and hence also the resonances in $\Sigma(\omega)$. However, as long as the rate is smaller relative to the resonance frequency, its effect is small.

The effect of temperature on the two experiments is indirect, through its effect on the intra- and inter-node scattering rates, as well as on the number of bulk filled Landau levels.
In all cases, the resulting effect scales like a power-law. This is in contrast to quantum oscillations, whose amplitude decreases exponentially with temperature \cite{Lifshitz}.

Finally, we comment on the applicability of these ideas to Dirac semimetals. Dirac SMs host similar inter-surface cyclotron orbits. In Dirac SMs, unlike in Weyl SMs, the counter propagating chiral LL reside at the same node. This reduces the robustness of the proposed effects, since now scatterings between the two chiral channels do not require large momentum transfer. Furthermore, in 
Cd$_3$As$_2$, the surface on which arcs are seen breaks the protecting symmetry \cite{QuantumOsc} and hence, one might worry that the arcs could be reconstructed into an ordinary Fermi surface.
Nonetheless, as long as the scattering rate between the different chiral levels is small, as suggested in \cite{WeylNonLocalTransport}, the effects should be visible also in Dirac SM. Note that the experimental observation of Fermi-arc orbits in Cd$_3$As$_2$ \cite{QuantumOscMeasur} is very encouraging in this regard.

\section{Estimates of Scales}
\begin{figure}[h!]
\centering
\includegraphics[width = \linewidth]{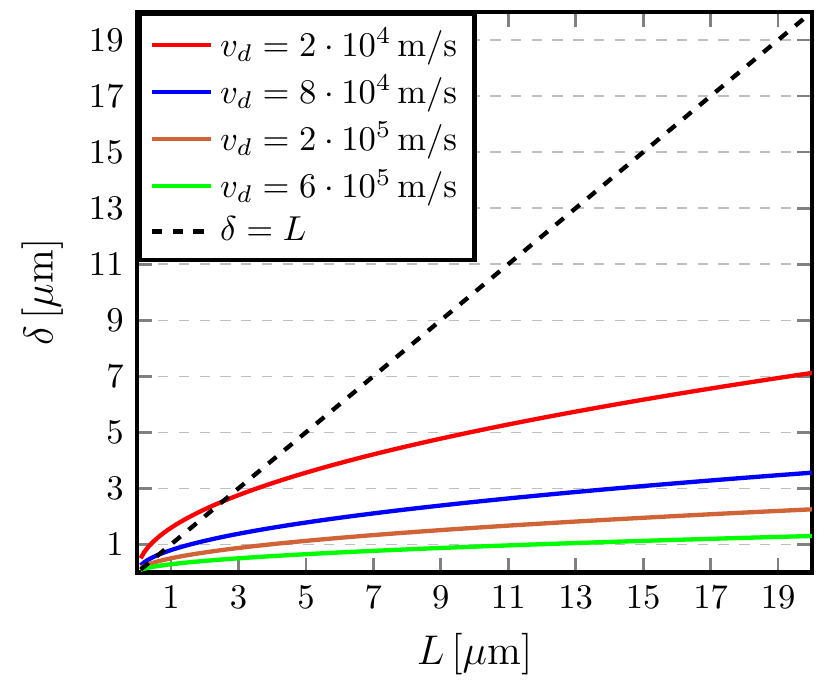}
\caption{Dependence of skin depth at the lowest resonance frequency, $\delta[\omega_c(L)]$, on sample thickness $L$ and for $\rho_0=10\,\mu\Omega \rm cm$; for $L\gtrsim 5\,\mu\text{m}$, $\delta\ll L$. \label{skinL}}
\end{figure}

We provide quantitative estimates of various relevant quantities. For typical values of $L=5-50\,\mu \rm m$, $v_0=1\cdot 10^5-6\cdot 10^5\,\rm m/s$ and $T_{\rm arc}< T_z$, the Weyl cyclotron frequency is $\omega_c=5-500\,\rm GHz$., i.e., in the microwave range.
In order for the described resonances to be a relevant probe, the sample should be thicker than the skin depth $\delta$ at moderate frequencies (if not, samples will be transparent at these frequencies independently of the magnetic field.) Assuming low-temperature scattering-rates similar to the ones measured in Dirac SM, we may extract a typical sample resistivity of the order $\sim10\,\mu\Omega \rm cm$ \cite{WeylExp1,WeylExp2}. This value produces a skin depth $\delta\approx 40\omega^{-1/2}\,\rm cm$, which is $\sim 1\,\mu\text{m}$ at microwave frequencies, serving as a lower bound on sample thickness. The upper bound is determined by the inter-node scattering processes: we require that $L\lesssim l_v$, the valley relaxation length, which can be tens of microns \cite{WeylNonLocalTransport}. As a final point, we remark on a subtlety: unlike usual cyclotron orbits, the frequency for electron motion on the nonlocal orbit connecting opposite surfaces is thickness-dependent. Since the Weyl cyclotron resonances occur at the harmonics of this frequency, in order to observe the resonant transmission, we require that the skin depth at the lowest resonant frequency is smaller than the sample size, $L$, which induces an implicit dependence of the skin depth on the thickness, $\delta = \delta[\omega_c(L)]$ in order to observe Weyl cyclotron resonance. Thus, we have an additional  constraint on the thickness, $\delta[\omega_c(L)]< L$. Fig.~\ref{skinL} shows the dependence of $\delta$ on the thickness for various values of the drift velocity $v_d$; we see that for $L\gtrsim 5\,\mu\text{m}$, this condition is indeed satisfied.
 
\section{Conclusions}
We have shown that inter-surface cyclotron orbits affect the electronic properties of Weyl semimetals already at the semi-classical level, and we proposed two experiments which directly probe these orbits. We demonstrated that, in the presence of a magnetic field, the transport in Weyl SMs is characterized by a non local conductivity which leads to the appearance of a magnetic field dependent non-local DC voltage and to a resonant transmission of electromagnetic radiation through macroscopic samples of Weyl semimetals.
Furthermore, we argued that the semi-classical origin of the effects reduces dramatically the experimental requirements of thin-films, low temperatures and pure samples.

\begin{acknowledgments}
S.A.P. thanks A. Vishwanath, D.A. Pesin, D.A. Abanin, N.P. Ong and especially A.C. Potter for illuminating discussions on transport properties of topological semimetals, and acknowledges support from NSF Grant DMR-1455366. Y.B and A.S. acknowledge support from the European Research Council under the European Unions Seventh Framework Programme (FP7/2007-2013) / ERC Project MUNATOP, Minerva foundation, and the U.S.-Israel BSF. E.B. was supported by the ISF under grant 1291/12, by the Minerva foundation, and by a Marie Curie CIG grant. S.A.P. and A.S. thank the UC Berkeley Richard B. Gump South Pacific Research Station in Moorea, French Polynesia, for hospitality during the 2014 International Workshop on Topological Phases and Quantum Computation.
\end{acknowledgments}

\appendix
\section{DC Setup}\label{app1}
\subsection{Solution of the Kirchoff-Ohm's equations}\label{app1a}
\begin{figure}[h]
\centering
\includegraphics[width = \linewidth]{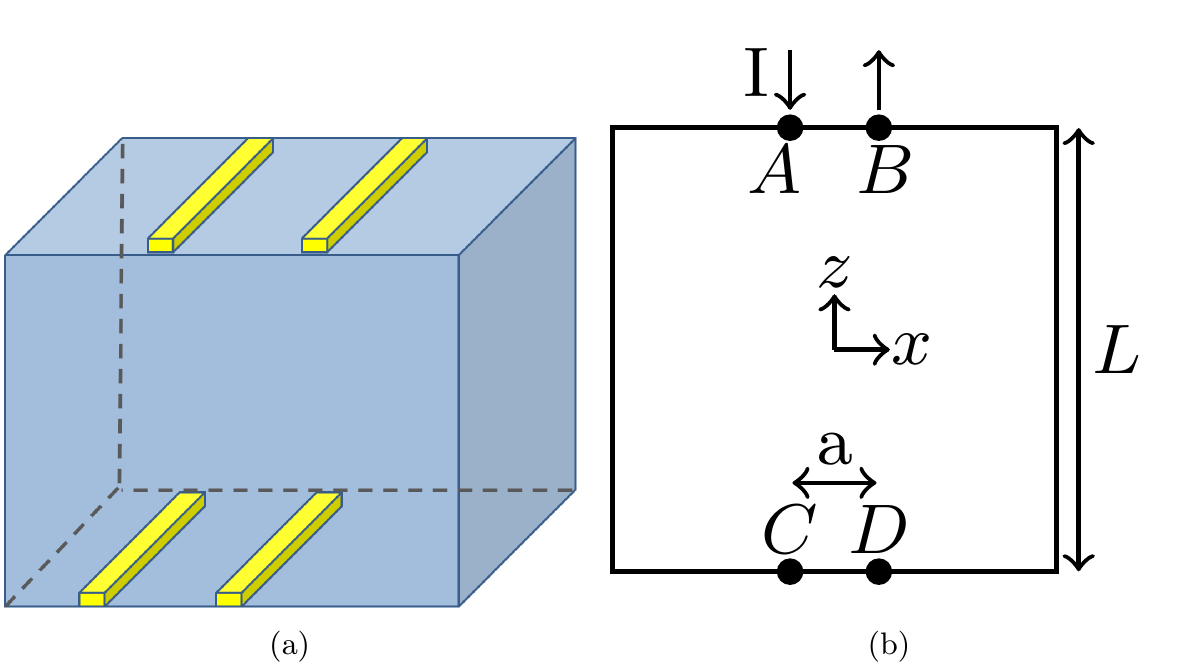}
\caption{(a) The DC setup. $y$ direction extent is much larger than $L$. (b) $x-z$ slice of the setup in (a). A current, $I$, is injected via contact $A$ and withdrawn from contact $B$. \label{fig_app1}}
\end{figure}

We consider a sample with four stripe-like Ohmic contacts lying along the $y$ direction and at a distance $a\ll L$ from one another as shown in \refFig{fig_app1}a. Two contacts are positioned on the $z=0$ plane and two on the $z=L$ plane. We assume that the $y$ direction extent is much larger than $L$, such that we expect translation invariance along $y$. Therefore, we consider a slice of the $x-z$ plane, \refFig{fig_app1}b. A current, $I$, is injected via contact $A$ and withdrawn from contact $B$. The voltage difference $V_{CD}$ will then be calculated.

The equations governing the current flow are:
\ba \label{kirch_app1}
&\bl{\nabla}\cdot\bl{J}=0 \\ \label{kirch_app2}
&\bl{\nabla}\times\bl{E}=0 \\ \label{kirch_app3}
&\bl{E}=\Sigma^{-1}\bl{J},
\end{align}
We choose to work here with the resistivity and not the conductivity since it simplifies the implementation of boundary conditions in this setup.
The generalized Ohm's law can be expressed in a matrix form: $J(z)=\int dz'\Sigma(z,z')E(z')$, where $\Sigma(z,z')=\sigma_0\delta_{z,z'}+\sigma_W(z,z')$.
Hence, the inverse relation is given by:
\begin{align} \nonumber
E(z)&=\Sigma^{-1}(z,z')J(z')=[\sigma_0\delta_{z,z'}+\sigma_W(z,z')]^{-1}J(z')\\ \label{smlimit}
&\approx \frac{1}{\sigma_0}J(z)-\frac{\sigma_W(z,z')}{\sigma_0^2}J(z')
\end{align}
In the last equality we assumed that $\sigma_W\ll\sigma_0$. This approximation is not essential. However, it simplifies the calculation. Later, we also discuss the opposite limit $\sigma_W\gg\sigma_0$. Using the form of the non-local conductivity, \refEq{nl}, we conclude that:
\begin{align}\nonumber
&\bl{E}^{\rm L}(z)=\frac{1}{\sigma_0}\bl{J}(z)\equiv\rho_0\bl{J}(z)\\ \nonumber
&\bl{E}^{\rm NL}(z)\approx -d\rho_{W}\delta(z-L)\left[\bl{J}(0)-\bl{J}(L)\right]+(0\leftrightarrow L),
\end{align}
where the L and NL refer to local and non-local, respectively, and $\rho_{W}=\sigma_W(\omega=0)/\sigma_0^2$. Notice that $\rho_W/\rho_0=\sigma_W/\sigma_0$.

Introducing the stream function $\psi(x,z)$ as follows:
\be
\bl{J}=\bl{\hat{y}}\times {\nabla}\psi\,\to\, J_x=-\partial_z\psi,\, J_z=\partial_x\psi.
\ee
\refEq{kirch_app1} and the $x$, $z$ components of \refEq{kirch_app2} are automatically satisfied. The $y$ components of \refEq{kirch_app2}:
\begin{align} \label{main_eq_app}
0&=(\bl{\nabla}\times\bl{E})_y=(\bl{\nabla}\times\bl{E}^{\rm L})_y+(\bl{\nabla}\times\bl{E}^{\rm NL})_y\\ \nonumber
&=\rho_0\nabla^2\psi+(\bl{\nabla}\times\bl{E}^{\rm NL})_y\,\to\,\rho_0\nabla^2\psi=-(\bl{\nabla}\times\bl{E}^{\rm NL})_y
\end{align}
Since no current can leave or enter the system except at the contacts, the stream
function $\psi$ must be constant along the boundaries. Since a current $I$ is injected at contact $A$ and withdrawn
at contact $B$, integrating $J_x=-\partial_z\psi$ across the step discontinuity at either contact implies that the jump in $\psi$ across contact $A\, (B)$ is $I\, (-I)$.
We choose $I=1$, thus, the boundary conditions are: $\psi=1$ on the boundary segment between $A$ and $B$, and $\psi=0$ on the all other boundary segments.
Notice that the current density $J$ lies along contours of constant $\psi$, with magnitude $|\nabla\psi|$.
Additionally, the voltage difference between any two points on the boundary is given by:
\be \label{volt}
V_{ij}=\int\limits_{x_i}^{x_j} dx E_x=\int\limits_{x_i}^{x_j} dx \rho_0J_x=-\int\limits_{x_i}^{x_j} dx \rho_0\partial_z\psi
\ee
For $\bl{E}^{\rm NL}=0$ the solution is depicted in \refFig{fig_app2}a. Evidently, the current
path extends a distance of the order of $a$ into the bulk, and is therefore vanishingly small at the $z = 0$ surface.
Inserting the solution in \refEq{volt} gives:
$V_{CD}/V_{AB}\approx 0.005.$

Next, we introduce the non local part $E^{\rm NL}$.
The boundary conditions remain unchanged. The solution to \refEq{main_eq_app} with $\rho_W/\rho_0=\sigma_W/\sigma_0=0.15$ is depicted in \refFig{fig_app2}b, and the voltage ratio becomes:
$V_{CD}/V_{AB}\approx -0.273.$

\begin{figure} [b]
\centering
\includegraphics[width = \linewidth]{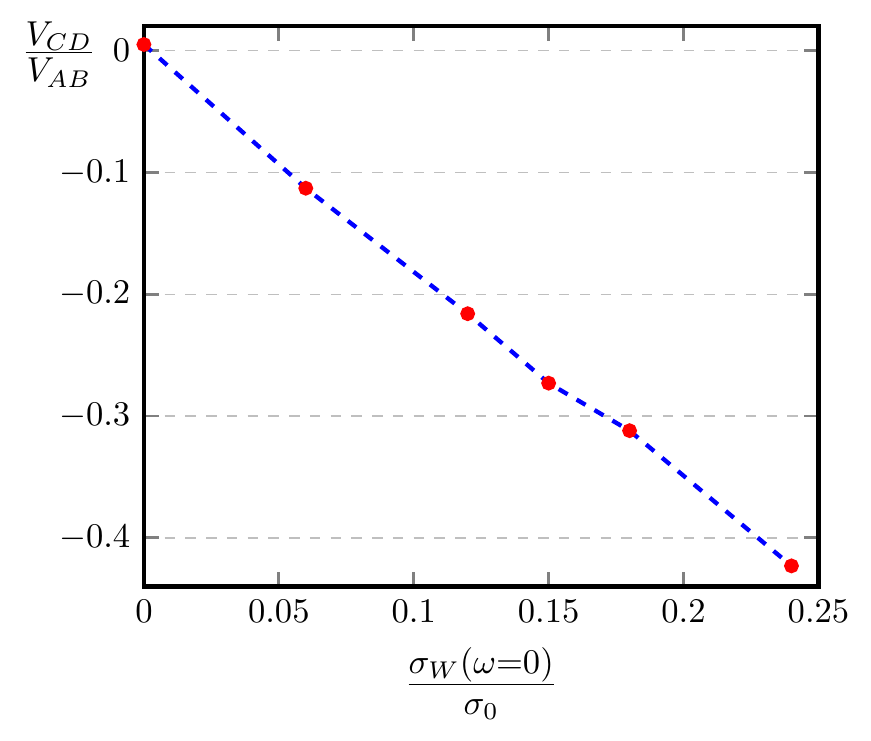}
\caption{The ratio of the voltages on the bottom and top surfaces as a function of the ratio $\sigma_W/\sigma_0$. The voltage ratio is calculated numerically, in the limit $\sigma_W/\sigma_0\ll1$, using the approximated Eqs.~\ref{kirch_app1}--\ref{kirch_app3}. In the limit $\sigma_W\gg\sigma_0$, the voltage ratio tends to unity. \label{DC_scaling}}
\end{figure}

In the limit $\sigma_W(\omega=0)\ll\sigma_0$, in which the voltage ratio may be calculated numerically, it is proportional to $\sigma_W/\sigma_0$. The voltage ratio is depicted in \refFig{DC_scaling}.

Next, we comment on the limit $\sigma_W\gg\sigma_0$. Neglecting the dependence of the current on $x$, we may write a one-dimensional model for the current in the $z$-direction. By discretizing the z-coordinate $z=(0,\Delta z,2\Delta z,\cdots,L)$, where $\Delta z$ is larger than the thickness of the Fermi arcs, the conductivity may be expressed as a finite matrix of the form: 
\be \Sigma_{z,z'}=
\begin{bmatrix}
 \sigma_0+\sigma_W & 0 & \cdots & 0 & -\sigma_W \\
 0 &  \sigma_0    &  &  0 & 0 \\
\vdots &   &  \ddots &   &  \vdots\\
0 & 0 &  & \sigma_0 & 0 \\
 -\sigma_W &  0 & \cdots &  0 & \sigma_0+\sigma_W
\end{bmatrix}.
\ee
Inverting the conductivity matrix yields the resistivity matrix:
\be \Sigma^{-1}=
\begin{bmatrix}
 A & 0 & \cdots & 0 & B \\
 0 &  1/\sigma_0  &  &  0 & 0 \\
\vdots &   &  \ddots &   &  \vdots\\
0 & 0 &  & 1/\sigma_0 & 0 \\
 B &  0 & \cdots &  0 & A 
\end{bmatrix},
\ee
where $A=\frac{1}{2\sigma_0}+\frac{1}{2(\sigma_0+2\sigma_W)}$ and $B=\frac{\sigma_W}{\sigma_0(\sigma_0+2\sigma_W)}$.

In agreement with \refEq{smlimit}, for $\sigma_W\ll\sigma_0$ the coefficients become $A\approx1/\sigma_0$ and $B\approx\sigma_W/\sigma_0^2$.
In the opposite limit, $\sigma_W\gg\sigma_0$, the coefficients become $A\approx B\approx1/(2\sigma_0)$. Independently of the current density profile, the condition $A\approx B$ forces the electric fields on the top and bottom surfaces to be equal, i.e., the voltage ratio tends to unity in the limit $\sigma_W\gg\sigma_0$. 

Finally, we comment on the locality of $\Sigma$ in the $x-y$ plane. In the above calculations, we assumed that $\Sigma$ is local in the $x-y$ plane. The locality breaks for length-scales smaller than $k_0\ell_B^2$. Nonetheless, as long as $k_0\ell_B^2\ll a$ the effects of the non-locality in the $x-y$ plane are negligible and the locality assumption is justified.

\section{AC setup --- Calculation of the transmission coefficient} \label{app2}
The Maxwell equations in SI units are:
\ba \label{maxwell}
&\bl{\nabla}\times\bl{E}=-\mu\frac{\partial \bl{H}}{\partial t}=\I\omega\mu\bl{H} \\
&\bl{\nabla}\times\bl{H}=\bl{J}+\epsilon\frac{\partial \bl{E}}{\partial t}=\bl{J}-\I\omega\epsilon\bl{E} \nonumber
\end{align}
where $\mu=\mu_{\rm 0}\mu_{\rm r}$ and $\epsilon=\epsilon_{\rm 0}\epsilon_{\rm r}$ are the permeability and permittivity of the metal.
We choose the applied a.c.~electric field to be along the $x$ direction. Hence, we may consider only the $x$ component of \refEq{maxwell}. The current in \refEq{maxwell} has two parts: $\bl{J}(z)=\sigma_{\rm 0}\bl{E}(z)+\bl{J}^{\rm NL}(z)$, where $\sigma_{\rm 0}$ is the conductivity, and $\bl{J}^{\rm NL}(z)$ is non-local part (see Eq.~5 in the main text). In general, the $\sigma_0$ depends on $\omega$. However, for microwave frequencies $\omega\tau \ll 1$, hence, we neglect the frequency dependence of $\sigma_{\rm 0}$.

\begin{figure}[t!]
\centering
\includegraphics[width = \linewidth]{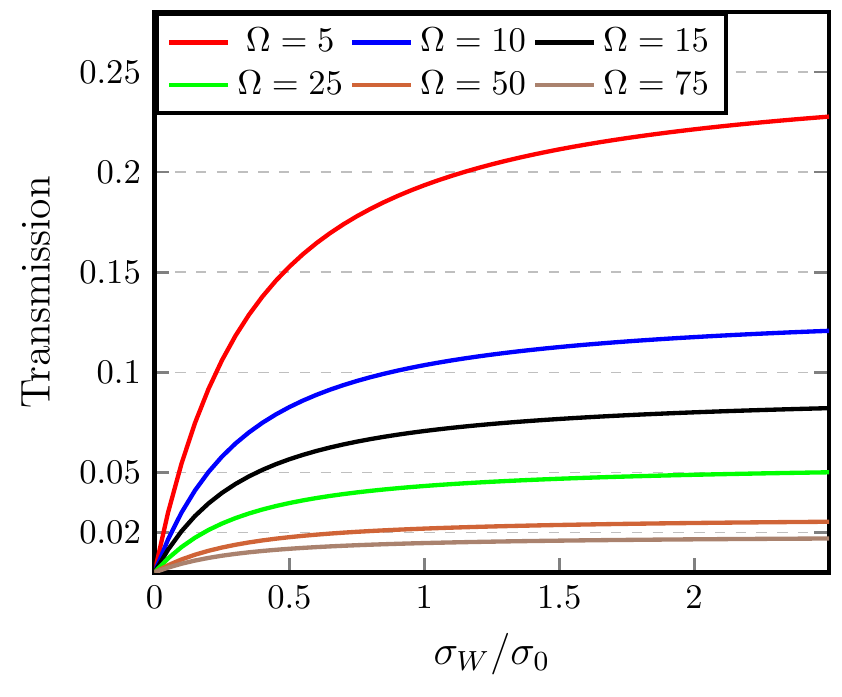}
\caption{The transmission coefficient as a function of $S=\sigma_{W}/\sigma_0$ for different values of $\Omega=(k\delta)^{-1}$ and for $\Delta=L/\delta=20$. \label{trans}}
\end{figure}

Inserting the two parts of $\bl{J}$ to \refEq{maxwell}, assuming translational invariance in the $xy$ plane, and introducing the dielectric function, $\varepsilon(\omega)=\epsilon_r+\frac{\I\sigma_0}{\omega\epsilon_0}$, \refEq{maxwell} becomes,
\be \label{max4}
\left[\partial_{zz}+\left(\frac{\omega}{c}\right)^2\varepsilon\right]E(z)=\I\omega\mu \bl{J}^{\rm NL}(z),
\ee
where we assumed $\mu_r\approx 1$ and employed the relation $\mu_0\epsilon_0=c^{-2}$.
Integrating \refEq{max4} form $0^-$ to $0^+$ and from $L^-$ to $L^+$ we find the boundary conditions (BC): $E_x(z)$ is a continuous function, in particular $E_x(0^-)=E_x(0^+)$, $E_x(L^-)=E_x(L^+)$, and
\begin{align}\nonumber
&(\partial_zE_x)_{z=0^+}-(\partial_zE_x)_{z=0^-}=-\I\omega \mu d\sigma_W\left[E_x(0)-E_x(L)\right], \\ \nonumber
&(\partial_zE_x)_{z=L^+}-(\partial_zE_x)_{z=L^-}=\I\omega \mu d\sigma_W\left[E_x(0)-E_x(L)\right].
\end{align}

The last two BC are equivalent to the requirement that the discontinuity of the tangential magnetic field is proportional to the surface current.
Next, we consider an incoming monochromatic field with a unit amplitude $E_0=1$. For $z\neq 0\,\mbox{or }L$, the solution to \refEq{max4} has the following general form:
\be
E_x(z)=\begin{cases} e^{-\I kz}+E_re^{\I kz} &\mbox{, } L<z \\
                     E_1e^{\I qz}+E_2e^{-\I qz} &\mbox{, } 0<z<L \\
										 E_te^{-\I kz} &\mbox{, } z<L\\
\end{cases},
\ee
where $k=\frac{\omega}{c}$ and $q=\frac{\omega}{c}\sqrt{\varepsilon(\omega)}$. In typical metals and for microwave frequencies $\sigma_0\gg\epsilon_0\omega$. In this regime, all the physical quantities may be recast in terms of following three dimensionless quantities: $\Omega=(k\delta)^{-1}$, $\Delta=L/\delta$ and $S=\sigma_{W}/\sigma_{0}$.
Imposing the BC, we find
\be\label{sys_dless_app}
\hat{M}\bmat E_1 \\ E_2 \\ E_r \\ E_t \emat=\bmat 1 \\ 1-2S\Omega \\ 0 \\ 2S\Omega \emat,
\ee
where the matrix $\hat{M}$ is given by
\begin{equation}\label{sys_matrix} \nonumber
\begin{psmallmatrix} 1 & 1 &-1 & 0 \\
       (1+\I)\Omega & -(1+\I)\Omega & 1+2S\Omega & -2S\Omega e^{\I \Delta/\Omega}\\
			e^{(\I-1)\Delta} & e^{-(\I-1)\Delta} & 0 & -e^{\I \Delta/\Omega}\\
			-(1+\I)\Omega e^{(\I-1)\Delta} & (1+\I)\Omega e^{-(\I-1)\Delta} & -2S\Omega & (2S\Omega+1)e^{\I \Delta/\Omega} \end{psmallmatrix}
\end{equation}
\refEq{sys_dless_app} may be solved for $|E_t|$ to yield the transmission coefficient as depicted in \refFig{trans}.

For $S=0$, the field in the metal decays as $E_se^{-z/\delta}$, where $\delta\sim (\omega\sigma_0)^{-1/2}$ is the skin-depth and the quantity $E_s$ is the field on the surface.
The ratio between the field on the surface and amplitude of the applied field is proportional to $(\omega\sigma_0)^{1/2}$. In particular, the dissipation is $\int{dz \mathbf{E}\cdot \mathbf{J}}\sim\delta\sigma_0E_s^2\propto(\omega/\sigma_0)^{1/2}$.
For a fixed $\sigma_0$, as $\omega$ increases, the decay becomes stronger, however, the value from which it decays becomes larger.

In the Weyl case, the mechanism we described transforms part of the absorption into transmission, with an effectiveness that increases with $\sigma_W/\sigma_0$. However, this mechanism does not eliminate the reflection, and hence the saturation of the transmission at a value that is smaller than one. The increase of absorption with $\omega$ in standard metals is transformed here into higher transmission.


\begin{thebibliography}{36}%
\makeatletter
\providecommand \@ifxundefined [1]{%
 \@ifx{#1\undefined}
}%
\providecommand \@ifnum [1]{%
 \ifnum #1\expandafter \@firstoftwo
 \else \expandafter \@secondoftwo
 \fi
}%
\providecommand \@ifx [1]{%
 \ifx #1\expandafter \@firstoftwo
 \else \expandafter \@secondoftwo
 \fi
}%
\providecommand \natexlab [1]{#1}%
\providecommand \enquote  [1]{``#1''}%
\providecommand \bibnamefont  [1]{#1}%
\providecommand \bibfnamefont [1]{#1}%
\providecommand \citenamefont [1]{#1}%
\providecommand \href@noop [0]{\@secondoftwo}%
\providecommand \href [0]{\begingroup \@sanitize@url \@href}%
\providecommand \@href[1]{\@@startlink{#1}\@@href}%
\providecommand \@@href[1]{\endgroup#1\@@endlink}%
\providecommand \@sanitize@url [0]{\catcode `\\12\catcode `\$12\catcode
  `\&12\catcode `\#12\catcode `\^12\catcode `\_12\catcode `\%12\relax}%
\providecommand \@@startlink[1]{}%
\providecommand \@@endlink[0]{}%
\providecommand \url  [0]{\begingroup\@sanitize@url \@url }%
\providecommand \@url [1]{\endgroup\@href {#1}{\urlprefix }}%
\providecommand \urlprefix  [0]{URL }%
\providecommand \Eprint [0]{\href }%
\providecommand \doibase [0]{http://dx.doi.org/}%
\providecommand \selectlanguage [0]{\@gobble}%
\providecommand \bibinfo  [0]{\@secondoftwo}%
\providecommand \bibfield  [0]{\@secondoftwo}%
\providecommand \translation [1]{[#1]}%
\providecommand \BibitemOpen [0]{}%
\providecommand \bibitemStop [0]{}%
\providecommand \bibitemNoStop [0]{.\EOS\space}%
\providecommand \EOS [0]{\spacefactor3000\relax}%
\providecommand \BibitemShut  [1]{\csname bibitem#1\endcsname}%
\let\auto@bib@innerbib\@empty
\bibitem [{\citenamefont {Hasan}\ and\ \citenamefont {Kane}(2010)}]{Top1}%
  \BibitemOpen
  \bibfield  {author} {\bibinfo {author} {\bibfnamefont {M.~Z.}\ \bibnamefont
  {Hasan}}\ and\ \bibinfo {author} {\bibfnamefont {C.~L.}\ \bibnamefont
  {Kane}},\ }\bibfield  {title} {\enquote {\bibinfo {title}
  {\textit{Colloquium} : Topological insulators},}\ }\href {\doibase
  10.1103/RevModPhys.82.3045} {\bibfield  {journal} {\bibinfo  {journal} {Rev.
  Mod. Phys.}\ }\textbf {\bibinfo {volume} {82}},\ \bibinfo {pages}
  {3045--3067} (\bibinfo {year} {2010})}\BibitemShut {NoStop}%
\bibitem [{\citenamefont {Qi}\ and\ \citenamefont {Zhang}(2011)}]{Top2}%
  \BibitemOpen
  \bibfield  {author} {\bibinfo {author} {\bibfnamefont {Xiao-Liang}\
  \bibnamefont {Qi}}\ and\ \bibinfo {author} {\bibfnamefont {Shou-Cheng}\
  \bibnamefont {Zhang}},\ }\bibfield  {title} {\enquote {\bibinfo {title}
  {Topological insulators and superconductors},}\ }\href {\doibase
  10.1103/RevModPhys.83.1057} {\bibfield  {journal} {\bibinfo  {journal} {Rev.
  Mod. Phys.}\ }\textbf {\bibinfo {volume} {83}},\ \bibinfo {pages}
  {1057--1110} (\bibinfo {year} {2011})}\BibitemShut {NoStop}%
\bibitem [{\citenamefont {{Kitaev}}(2009)}]{Kitaev}%
  \BibitemOpen
  \bibfield  {author} {\bibinfo {author} {\bibfnamefont {A.}~\bibnamefont
  {{Kitaev}}},\ }\bibfield  {title} {\enquote {\bibinfo {title} {{Periodic
  table for topological insulators and superconductors}},}\ }in\ \href
  {\doibase 10.1063/1.3149495} {\emph {\bibinfo {booktitle} {American Institute
  of Physics Conference Series}}},\ \bibinfo {series} {American Institute of
  Physics Conference Series}, Vol.\ \bibinfo {volume} {1134},\ \bibinfo
  {editor} {edited by\ \bibinfo {editor} {\bibfnamefont {V.}~\bibnamefont
  {{Lebedev}}}\ and\ \bibinfo {editor} {\bibfnamefont {M.}~\bibnamefont
  {{Feigel'man}}}}\ (\bibinfo {year} {2009})\ pp.\ \bibinfo {pages}
  {22--30}\BibitemShut {NoStop}%
\bibitem [{\citenamefont {Schnyder}\ \emph {et~al.}(2008)\citenamefont
  {Schnyder}, \citenamefont {Ryu}, \citenamefont {Furusaki},\ and\
  \citenamefont {Ludwig}}]{class}%
  \BibitemOpen
  \bibfield  {author} {\bibinfo {author} {\bibfnamefont {Andreas~P.}\
  \bibnamefont {Schnyder}}, \bibinfo {author} {\bibfnamefont {Shinsei}\
  \bibnamefont {Ryu}}, \bibinfo {author} {\bibfnamefont {Akira}\ \bibnamefont
  {Furusaki}}, \ and\ \bibinfo {author} {\bibfnamefont {Andreas W.~W.}\
  \bibnamefont {Ludwig}},\ }\bibfield  {title} {\enquote {\bibinfo {title}
  {Classification of topological insulators and superconductors in three
  spatial dimensions},}\ }\href {\doibase 10.1103/PhysRevB.78.195125}
  {\bibfield  {journal} {\bibinfo  {journal} {Phys. Rev. B}\ }\textbf {\bibinfo
  {volume} {78}},\ \bibinfo {pages} {195125} (\bibinfo {year}
  {2008})}\BibitemShut {NoStop}%
\bibitem [{\citenamefont {Matsuura}\ \emph {et~al.}(2013)\citenamefont
  {Matsuura}, \citenamefont {Chang}, \citenamefont {Schnyder},\ and\
  \citenamefont {Ryu}}]{gapless1}%
  \BibitemOpen
  \bibfield  {author} {\bibinfo {author} {\bibfnamefont {Shunji}\ \bibnamefont
  {Matsuura}}, \bibinfo {author} {\bibfnamefont {Po-Yao}\ \bibnamefont
  {Chang}}, \bibinfo {author} {\bibfnamefont {Andreas~P}\ \bibnamefont
  {Schnyder}}, \ and\ \bibinfo {author} {\bibfnamefont {Shinsei}\ \bibnamefont
  {Ryu}},\ }\bibfield  {title} {\enquote {\bibinfo {title} {Protected boundary
  states in gapless topological phases},}\ }\href
  {http://stacks.iop.org/1367-2630/15/i=6/a=065001} {\bibfield  {journal}
  {\bibinfo  {journal} {New Journal of Physics}\ }\textbf {\bibinfo {volume}
  {15}},\ \bibinfo {pages} {065001} (\bibinfo {year} {2013})}\BibitemShut
  {NoStop}%
\bibitem [{\citenamefont {Baum}\ \emph {et~al.}(2015)\citenamefont {Baum},
  \citenamefont {Posske}, \citenamefont {Fulga}, \citenamefont {Trauzettel},\
  and\ \citenamefont {Stern}}]{gapless2}%
  \BibitemOpen
  \bibfield  {author} {\bibinfo {author} {\bibfnamefont {Yuval}\ \bibnamefont
  {Baum}}, \bibinfo {author} {\bibfnamefont {Thore}\ \bibnamefont {Posske}},
  \bibinfo {author} {\bibfnamefont {Ion~Cosma}\ \bibnamefont {Fulga}}, \bibinfo
  {author} {\bibfnamefont {Bj\"orn}\ \bibnamefont {Trauzettel}}, \ and\
  \bibinfo {author} {\bibfnamefont {Ady}\ \bibnamefont {Stern}},\ }\bibfield
  {title} {\enquote {\bibinfo {title} {Coexisting edge states and gapless bulk
  in topological states of matter},}\ }\href {\doibase
  10.1103/PhysRevLett.114.136801} {\bibfield  {journal} {\bibinfo  {journal}
  {Phys. Rev. Lett.}\ }\textbf {\bibinfo {volume} {114}},\ \bibinfo {pages}
  {136801} (\bibinfo {year} {2015})}\BibitemShut {NoStop}%
\bibitem [{\citenamefont {Keselman}\ and\ \citenamefont
  {Berg}(2015)}]{gapless3}%
  \BibitemOpen
  \bibfield  {author} {\bibinfo {author} {\bibfnamefont {Anna}\ \bibnamefont
  {Keselman}}\ and\ \bibinfo {author} {\bibfnamefont {Erez}\ \bibnamefont
  {Berg}},\ }\bibfield  {title} {\enquote {\bibinfo {title} {Gapless
  symmetry-protected topological phase of fermions in one dimension},}\ }\href
  {\doibase 10.1103/PhysRevB.91.235309} {\bibfield  {journal} {\bibinfo
  {journal} {Phys. Rev. B}\ }\textbf {\bibinfo {volume} {91}},\ \bibinfo
  {pages} {235309} (\bibinfo {year} {2015})}\BibitemShut {NoStop}%
\bibitem [{\citenamefont {{Turner}}\ and\ \citenamefont
  {{Vishwanath}}(2013)}]{WeylRevTurner}%
  \BibitemOpen
  \bibfield  {author} {\bibinfo {author} {\bibfnamefont {A.~M.}\ \bibnamefont
  {{Turner}}}\ and\ \bibinfo {author} {\bibfnamefont {A.}~\bibnamefont
  {{Vishwanath}}},\ }\bibfield  {title} {\enquote {\bibinfo {title} {{Beyond
  Band Insulators: Topology of Semi-metals and Interacting Phases}},}\
  }\href@noop {} {\bibfield  {journal} {\bibinfo  {journal} {ArXiv e-prints}\ }
  (\bibinfo {year} {2013})},\ \Eprint {http://arxiv.org/abs/1301.0330}
  {arXiv:1301.0330 [cond-mat.str-el]} \BibitemShut {NoStop}%
\bibitem [{\citenamefont {Wan}\ \emph {et~al.}(2011)\citenamefont {Wan},
  \citenamefont {Turner}, \citenamefont {Vishwanath},\ and\ \citenamefont
  {Savrasov}}]{WeylArcsWen}%
  \BibitemOpen
  \bibfield  {author} {\bibinfo {author} {\bibfnamefont {Xiangang}\
  \bibnamefont {Wan}}, \bibinfo {author} {\bibfnamefont {Ari~M.}\ \bibnamefont
  {Turner}}, \bibinfo {author} {\bibfnamefont {Ashvin}\ \bibnamefont
  {Vishwanath}}, \ and\ \bibinfo {author} {\bibfnamefont {Sergey~Y.}\
  \bibnamefont {Savrasov}},\ }\bibfield  {title} {\enquote {\bibinfo {title}
  {Topological semimetal and fermi-arc surface states in the electronic
  structure of pyrochlore iridates},}\ }\href {\doibase
  10.1103/PhysRevB.83.205101} {\bibfield  {journal} {\bibinfo  {journal} {Phys.
  Rev. B}\ }\textbf {\bibinfo {volume} {83}},\ \bibinfo {pages} {205101}
  (\bibinfo {year} {2011})}\BibitemShut {NoStop}%
\bibitem [{\citenamefont {Burkov}\ and\ \citenamefont
  {Balents}(2011)}]{WeylFromTI}%
  \BibitemOpen
  \bibfield  {author} {\bibinfo {author} {\bibfnamefont {A.~A.}\ \bibnamefont
  {Burkov}}\ and\ \bibinfo {author} {\bibfnamefont {Leon}\ \bibnamefont
  {Balents}},\ }\bibfield  {title} {\enquote {\bibinfo {title} {Weyl semimetal
  in a topological insulator multilayer},}\ }\href {\doibase
  10.1103/PhysRevLett.107.127205} {\bibfield  {journal} {\bibinfo  {journal}
  {Phys. Rev. Lett.}\ }\textbf {\bibinfo {volume} {107}},\ \bibinfo {pages}
  {127205} (\bibinfo {year} {2011})}\BibitemShut {NoStop}%
\bibitem [{\citenamefont {Hosur}\ and\ \citenamefont {Qi}(2013)}]{Hosur}%
  \BibitemOpen
  \bibfield  {author} {\bibinfo {author} {\bibfnamefont {Pavan}\ \bibnamefont
  {Hosur}}\ and\ \bibinfo {author} {\bibfnamefont {Xiaoliang}\ \bibnamefont
  {Qi}},\ }\bibfield  {title} {\enquote {\bibinfo {title} {Recent developments
  in transport phenomena in weyl semimetals},}\ }\href {\doibase
  http://dx.doi.org/10.1016/j.crhy.2013.10.010} {\bibfield  {journal} {\bibinfo
   {journal} {Comptes Rendus Physique}\ }\textbf {\bibinfo {volume} {14}},\
  \bibinfo {pages} {857 -- 870} (\bibinfo {year} {2013})}\BibitemShut {NoStop}%
\bibitem [{Note1()}]{Note1}%
  \BibitemOpen
  \bibinfo {note} {The general form of the Weyl Hamiltonian is $k_iA_{ij}\sigma
  _j+f(\protect \textbf {k})\sigma _0$, where $A$ is a $3\times 3$ matrix and
  $f$ is a linear function of $k$. Our results should not depend on the exact
  form of $A$ and $f$, at least qualitatively. For simplicity, we consider only
  the isotropic case.}\BibitemShut {Stop}%
\bibitem [{\citenamefont {Potter}\ \emph {et~al.}(2014)\citenamefont {Potter},
  \citenamefont {Kimchi},\ and\ \citenamefont {Vishwanath}}]{QuantumOsc}%
  \BibitemOpen
  \bibfield  {author} {\bibinfo {author} {\bibfnamefont {Andrew~C.}\
  \bibnamefont {Potter}}, \bibinfo {author} {\bibfnamefont {Itamar}\
  \bibnamefont {Kimchi}}, \ and\ \bibinfo {author} {\bibfnamefont {Ashvin}\
  \bibnamefont {Vishwanath}},\ }\bibfield  {title} {\enquote {\bibinfo {title}
  {Quantum oscillations from surface fermi arcs in weyl and dirac
  semimetals},}\ }\href {\doibase 10.1038/ncomms6161} {\bibfield  {journal}
  {\bibinfo  {journal} {Nature Communications}\ }\textbf {\bibinfo {volume}
  {5}} (\bibinfo {year} {2014}),\ 10.1038/ncomms6161}\BibitemShut {NoStop}%
\bibitem [{\citenamefont {{Moll}}\ \emph {et~al.}(2015)\citenamefont {{Moll}},
  \citenamefont {{Nair}}, \citenamefont {{Helm}}, \citenamefont {{Potter}},
  \citenamefont {{Kimchi}}, \citenamefont {{Vishwanath}},\ and\ \citenamefont
  {{Analytis}}}]{QuantumOscMeasur}%
  \BibitemOpen
  \bibfield  {author} {\bibinfo {author} {\bibfnamefont {P.~J.~W.}\
  \bibnamefont {{Moll}}}, \bibinfo {author} {\bibfnamefont {N.~L.}\
  \bibnamefont {{Nair}}}, \bibinfo {author} {\bibfnamefont {T.}~\bibnamefont
  {{Helm}}}, \bibinfo {author} {\bibfnamefont {A.~C.}\ \bibnamefont
  {{Potter}}}, \bibinfo {author} {\bibfnamefont {I.}~\bibnamefont {{Kimchi}}},
  \bibinfo {author} {\bibfnamefont {A.}~\bibnamefont {{Vishwanath}}}, \ and\
  \bibinfo {author} {\bibfnamefont {J.~G.}\ \bibnamefont {{Analytis}}},\
  }\bibfield  {title} {\enquote {\bibinfo {title} {{Chirality transfer dynamics
  in quantum orbits in the Dirac semi-metal Cd$_3$As$_2$}},}\ }\href@noop
  {} {\bibfield  {journal} {\bibinfo  {journal} {ArXiv e-prints}\ } (\bibinfo
  {year} {2015})},\ \Eprint {http://arxiv.org/abs/1505.02817} {arXiv:1505.02817
  [cond-mat.mes-hall]} \BibitemShut {NoStop}%
\bibitem [{\citenamefont {Son}\ and\ \citenamefont {Spivak}(2013)}]{Son2013}%
  \BibitemOpen
  \bibfield  {author} {\bibinfo {author} {\bibfnamefont {DT}~\bibnamefont
  {Son}}\ and\ \bibinfo {author} {\bibfnamefont {BZ}~\bibnamefont {Spivak}},\
  }\bibfield  {title} {\enquote {\bibinfo {title} {Chiral anomaly and classical
  negative magnetoresistance of weyl metals},}\ }\href@noop {} {\bibfield
  {journal} {\bibinfo  {journal} {Physical Review B}\ }\textbf {\bibinfo
  {volume} {88}},\ \bibinfo {pages} {104412} (\bibinfo {year}
  {2013})}\BibitemShut {NoStop}%
\bibitem [{\citenamefont {Parameswaran}\ \emph {et~al.}(2014)\citenamefont
  {Parameswaran}, \citenamefont {Grover}, \citenamefont {Abanin}, \citenamefont
  {Pesin},\ and\ \citenamefont {Vishwanath}}]{WeylNonLocalTransport}%
  \BibitemOpen
  \bibfield  {author} {\bibinfo {author} {\bibfnamefont {S.~A.}\ \bibnamefont
  {Parameswaran}}, \bibinfo {author} {\bibfnamefont {T.}~\bibnamefont
  {Grover}}, \bibinfo {author} {\bibfnamefont {D.~A.}\ \bibnamefont {Abanin}},
  \bibinfo {author} {\bibfnamefont {D.~A.}\ \bibnamefont {Pesin}}, \ and\
  \bibinfo {author} {\bibfnamefont {A.}~\bibnamefont {Vishwanath}},\ }\bibfield
   {title} {\enquote {\bibinfo {title} {Probing the chiral anomaly with
  nonlocal transport in three-dimensional topological semimetals},}\ }\href
  {\doibase 10.1103/PhysRevX.4.031035} {\bibfield  {journal} {\bibinfo
  {journal} {Phys. Rev. X}\ }\textbf {\bibinfo {volume} {4}},\ \bibinfo {pages}
  {031035} (\bibinfo {year} {2014})}\BibitemShut {NoStop}%
\bibitem [{\citenamefont {Witczak-Krempa}\ and\ \citenamefont
  {Kim}(2012)}]{WeylPyroc}%
  \BibitemOpen
  \bibfield  {author} {\bibinfo {author} {\bibfnamefont {William}\ \bibnamefont
  {Witczak-Krempa}}\ and\ \bibinfo {author} {\bibfnamefont {Yong~Baek}\
  \bibnamefont {Kim}},\ }\bibfield  {title} {\enquote {\bibinfo {title}
  {Topological and magnetic phases of interacting electrons in the pyrochlore
  iridates},}\ }\href {\doibase 10.1103/PhysRevB.85.045124} {\bibfield
  {journal} {\bibinfo  {journal} {Phys. Rev. B}\ }\textbf {\bibinfo {volume}
  {85}},\ \bibinfo {pages} {045124} (\bibinfo {year} {2012})}\BibitemShut
  {NoStop}%
\bibitem [{\citenamefont {Xu}\ \emph {et~al.}(2011)\citenamefont {Xu},
  \citenamefont {Weng}, \citenamefont {Wang}, \citenamefont {Dai},\ and\
  \citenamefont {Fang}}]{WeylMatHg}%
  \BibitemOpen
  \bibfield  {author} {\bibinfo {author} {\bibfnamefont {Gang}\ \bibnamefont
  {Xu}}, \bibinfo {author} {\bibfnamefont {Hongming}\ \bibnamefont {Weng}},
  \bibinfo {author} {\bibfnamefont {Zhijun}\ \bibnamefont {Wang}}, \bibinfo
  {author} {\bibfnamefont {Xi}~\bibnamefont {Dai}}, \ and\ \bibinfo {author}
  {\bibfnamefont {Zhong}\ \bibnamefont {Fang}},\ }\bibfield  {title} {\enquote
  {\bibinfo {title} {Chern semimetal and the quantized anomalous hall effect in
  ${\mathrm{HgCr}}_{2}{\mathrm{Se}}_{4}$},}\ }\href {\doibase
  10.1103/PhysRevLett.107.186806} {\bibfield  {journal} {\bibinfo  {journal}
  {Phys. Rev. Lett.}\ }\textbf {\bibinfo {volume} {107}},\ \bibinfo {pages}
  {186806} (\bibinfo {year} {2011})}\BibitemShut {NoStop}%
\bibitem [{\citenamefont {Weng}\ \emph {et~al.}(2015)\citenamefont {Weng},
  \citenamefont {Fang}, \citenamefont {Fang}, \citenamefont {Bernevig},\ and\
  \citenamefont {Dai}}]{tranmetal1}%
  \BibitemOpen
  \bibfield  {author} {\bibinfo {author} {\bibfnamefont {Hongming}\
  \bibnamefont {Weng}}, \bibinfo {author} {\bibfnamefont {Chen}\ \bibnamefont
  {Fang}}, \bibinfo {author} {\bibfnamefont {Zhong}\ \bibnamefont {Fang}},
  \bibinfo {author} {\bibfnamefont {B.~Andrei}\ \bibnamefont {Bernevig}}, \
  and\ \bibinfo {author} {\bibfnamefont {Xi}~\bibnamefont {Dai}},\ }\bibfield
  {title} {\enquote {\bibinfo {title} {Weyl semimetal phase in
  noncentrosymmetric transition-metal monophosphides},}\ }\href {\doibase
  10.1103/PhysRevX.5.011029} {\bibfield  {journal} {\bibinfo  {journal} {Phys.
  Rev. X}\ }\textbf {\bibinfo {volume} {5}},\ \bibinfo {pages} {011029}
  (\bibinfo {year} {2015})}\BibitemShut {NoStop}%
\bibitem [{\citenamefont {Huang}\ \emph {et~al.}(2015)\citenamefont {Huang}
  \emph {et~al.}}]{tranmetal2}%
  \BibitemOpen
  \bibfield  {author} {\bibinfo {author} {\bibfnamefont {Shin-Ming}\
  \bibnamefont {Huang}} \emph {et~al.},\ }\bibfield  {title} {\enquote
  {\bibinfo {title} {A weyl fermion semimetal with surface fermi arcs in the
  transition metal monopnictide taas class},}\ }\href {\doibase
  10.1038/ncomms8373} {\bibfield  {journal} {\bibinfo  {journal} {Nature
  Communications}\ }\textbf {\bibinfo {volume} {6}} (\bibinfo {year} {2015}),\
  10.1038/ncomms8373}\BibitemShut {NoStop}%
\bibitem [{\citenamefont {{Xu}}\ \emph {et~al.}(2015)\citenamefont {{Xu}},
  \citenamefont {{Belopolski}}, \citenamefont {{Alidoust}}, \citenamefont
  {{Neupane}}, \citenamefont {{Zhang}}, \citenamefont {{Sankar}}, \citenamefont
  {{Huang}}, \citenamefont {{Lee}}, \citenamefont {{Chang}}, \citenamefont
  {{Wang}}, \citenamefont {{Bian}}, \citenamefont {{Zheng}}, \citenamefont
  {{Sanchez}}, \citenamefont {{Chou}}, \citenamefont {{Lin}}, \citenamefont
  {{Jia}},\ and\ \citenamefont {{Zahid Hasan}}}]{TaAs}%
  \BibitemOpen
  \bibfield  {author} {\bibinfo {author} {\bibfnamefont {S.-Y.}\ \bibnamefont
  {{Xu}}}, \bibinfo {author} {\bibfnamefont {I.}~\bibnamefont {{Belopolski}}},
  \bibinfo {author} {\bibfnamefont {N.}~\bibnamefont {{Alidoust}}}, \bibinfo
  {author} {\bibfnamefont {M.}~\bibnamefont {{Neupane}}}, \bibinfo {author}
  {\bibfnamefont {C.}~\bibnamefont {{Zhang}}}, \bibinfo {author} {\bibfnamefont
  {R.}~\bibnamefont {{Sankar}}}, \bibinfo {author} {\bibfnamefont {S.-M.}\
  \bibnamefont {{Huang}}}, \bibinfo {author} {\bibfnamefont {C.-C.}\
  \bibnamefont {{Lee}}}, \bibinfo {author} {\bibfnamefont {G.}~\bibnamefont
  {{Chang}}}, \bibinfo {author} {\bibfnamefont {B.}~\bibnamefont {{Wang}}},
  \bibinfo {author} {\bibfnamefont {G.}~\bibnamefont {{Bian}}}, \bibinfo
  {author} {\bibfnamefont {H.}~\bibnamefont {{Zheng}}}, \bibinfo {author}
  {\bibfnamefont {D.~S.}\ \bibnamefont {{Sanchez}}}, \bibinfo {author}
  {\bibfnamefont {F.}~\bibnamefont {{Chou}}}, \bibinfo {author} {\bibfnamefont
  {H.}~\bibnamefont {{Lin}}}, \bibinfo {author} {\bibfnamefont
  {S.}~\bibnamefont {{Jia}}}, \ and\ \bibinfo {author} {\bibfnamefont
  {M.}~\bibnamefont {{Zahid Hasan}}},\ }\bibfield  {title} {\enquote {\bibinfo
  {title} {{Discovery of a Weyl Fermion Semimetal and Topological Fermi
  Arcs}},}\ }\href@noop {} {\bibfield  {journal} {\bibinfo  {journal} {ArXiv
  e-prints}\ } (\bibinfo {year} {2015})},\ \Eprint
  {http://arxiv.org/abs/1502.03807} {arXiv:1502.03807 [cond-mat.mes-hall]}
  \BibitemShut {NoStop}%
\bibitem [{\citenamefont {Lv}\ \emph {et~al.}(2015)\citenamefont {Lv},
  \citenamefont {Weng}, \citenamefont {Fu}, \citenamefont {Wang}, \citenamefont
  {Miao}, \citenamefont {Ma}, \citenamefont {Richard}, \citenamefont {Huang},
  \citenamefont {Zhao}, \citenamefont {Chen}, \citenamefont {Fang},
  \citenamefont {Dai}, \citenamefont {Qian},\ and\ \citenamefont
  {Ding}}]{TaAs2}%
  \BibitemOpen
  \bibfield  {author} {\bibinfo {author} {\bibfnamefont {B.~Q.}\ \bibnamefont
  {Lv}}, \bibinfo {author} {\bibfnamefont {H.~M.}\ \bibnamefont {Weng}},
  \bibinfo {author} {\bibfnamefont {B.~B.}\ \bibnamefont {Fu}}, \bibinfo
  {author} {\bibfnamefont {X.~P.}\ \bibnamefont {Wang}}, \bibinfo {author}
  {\bibfnamefont {H.}~\bibnamefont {Miao}}, \bibinfo {author} {\bibfnamefont
  {J.}~\bibnamefont {Ma}}, \bibinfo {author} {\bibfnamefont {P.}~\bibnamefont
  {Richard}}, \bibinfo {author} {\bibfnamefont {X.~C.}\ \bibnamefont {Huang}},
  \bibinfo {author} {\bibfnamefont {L.~X.}\ \bibnamefont {Zhao}}, \bibinfo
  {author} {\bibfnamefont {G.~F.}\ \bibnamefont {Chen}}, \bibinfo {author}
  {\bibfnamefont {Z.}~\bibnamefont {Fang}}, \bibinfo {author} {\bibfnamefont
  {X.}~\bibnamefont {Dai}}, \bibinfo {author} {\bibfnamefont {T.}~\bibnamefont
  {Qian}}, \ and\ \bibinfo {author} {\bibfnamefont {H.}~\bibnamefont {Ding}},\
  }\bibfield  {title} {\enquote {\bibinfo {title} {Experimental discovery of
  weyl semimetal taas},}\ }\href {\doibase 10.1103/PhysRevX.5.031013}
  {\bibfield  {journal} {\bibinfo  {journal} {Phys. Rev. X}\ }\textbf {\bibinfo
  {volume} {5}},\ \bibinfo {pages} {031013} (\bibinfo {year}
  {2015})}\BibitemShut {NoStop}%
\bibitem [{\citenamefont {{Lv}}\ \emph {et~al.}(2015)\citenamefont {{Lv}},
  \citenamefont {{Xu}}, \citenamefont {{Weng}}, \citenamefont {{Ma}},
  \citenamefont {{Richard}}, \citenamefont {{Huang}}, \citenamefont {{Zhao}},
  \citenamefont {{Chen}}, \citenamefont {{Matt}}, \citenamefont {{Bisti}},
  \citenamefont {{Strokov}}, \citenamefont {{Mesot}}, \citenamefont {{Fang}},
  \citenamefont {{Dai}}, \citenamefont {{Qian}}, \citenamefont {{Shi}},\ and\
  \citenamefont {{Ding}}}]{TaAs3}%
  \BibitemOpen
  \bibfield  {author} {\bibinfo {author} {\bibfnamefont {B.~Q.}\ \bibnamefont
  {{Lv}}}, \bibinfo {author} {\bibfnamefont {N.}~\bibnamefont {{Xu}}}, \bibinfo
  {author} {\bibfnamefont {H.~M.}\ \bibnamefont {{Weng}}}, \bibinfo {author}
  {\bibfnamefont {J.~Z.}\ \bibnamefont {{Ma}}}, \bibinfo {author}
  {\bibfnamefont {P.}~\bibnamefont {{Richard}}}, \bibinfo {author}
  {\bibfnamefont {X.~C.}\ \bibnamefont {{Huang}}}, \bibinfo {author}
  {\bibfnamefont {L.~X.}\ \bibnamefont {{Zhao}}}, \bibinfo {author}
  {\bibfnamefont {G.~F.}\ \bibnamefont {{Chen}}}, \bibinfo {author}
  {\bibfnamefont {C.}~\bibnamefont {{Matt}}}, \bibinfo {author} {\bibfnamefont
  {F.}~\bibnamefont {{Bisti}}}, \bibinfo {author} {\bibfnamefont
  {V.}~\bibnamefont {{Strokov}}}, \bibinfo {author} {\bibfnamefont
  {J.}~\bibnamefont {{Mesot}}}, \bibinfo {author} {\bibfnamefont
  {Z.}~\bibnamefont {{Fang}}}, \bibinfo {author} {\bibfnamefont
  {X.}~\bibnamefont {{Dai}}}, \bibinfo {author} {\bibfnamefont
  {T.}~\bibnamefont {{Qian}}}, \bibinfo {author} {\bibfnamefont
  {M.}~\bibnamefont {{Shi}}}, \ and\ \bibinfo {author} {\bibfnamefont
  {H.}~\bibnamefont {{Ding}}},\ }\bibfield  {title} {\enquote {\bibinfo {title}
  {{Observation of Weyl nodes in TaAs}},}\ }\href@noop {} {\bibfield  {journal}
  {\bibinfo  {journal} {ArXiv e-prints}\ } (\bibinfo {year} {2015})},\ \Eprint
  {http://arxiv.org/abs/1503.09188} {arXiv:1503.09188 [cond-mat.mtrl-sci]}
  \BibitemShut {NoStop}%
\bibitem [{\citenamefont {{Zhang}}\ \emph {et~al.}(2015)\citenamefont
  {{Zhang}}, \citenamefont {{Yuan}}, \citenamefont {{Xu}}, \citenamefont
  {{Lin}}, \citenamefont {{Tong}}, \citenamefont {{Zahid Hasan}}, \citenamefont
  {{Wang}}, \citenamefont {{Zhang}},\ and\ \citenamefont {{Jia}}}]{TaAs4}%
  \BibitemOpen
  \bibfield  {author} {\bibinfo {author} {\bibfnamefont {C.}~\bibnamefont
  {{Zhang}}}, \bibinfo {author} {\bibfnamefont {Z.}~\bibnamefont {{Yuan}}},
  \bibinfo {author} {\bibfnamefont {S.}~\bibnamefont {{Xu}}}, \bibinfo {author}
  {\bibfnamefont {Z.}~\bibnamefont {{Lin}}}, \bibinfo {author} {\bibfnamefont
  {B.}~\bibnamefont {{Tong}}}, \bibinfo {author} {\bibfnamefont
  {M.}~\bibnamefont {{Zahid Hasan}}}, \bibinfo {author} {\bibfnamefont
  {J.}~\bibnamefont {{Wang}}}, \bibinfo {author} {\bibfnamefont
  {C.}~\bibnamefont {{Zhang}}}, \ and\ \bibinfo {author} {\bibfnamefont
  {S.}~\bibnamefont {{Jia}}},\ }\bibfield  {title} {\enquote {\bibinfo {title}
  {{Tantalum Monoarsenide: an Exotic Compensated Semimetal}},}\ }\href@noop {}
  {\bibfield  {journal} {\bibinfo  {journal} {ArXiv e-prints}\ } (\bibinfo
  {year} {2015})},\ \Eprint {http://arxiv.org/abs/1502.00251} {arXiv:1502.00251
  [cond-mat.mtrl-sci]} \BibitemShut {NoStop}%
\bibitem [{\citenamefont {{Huang}}\ \emph {et~al.}(2015)\citenamefont
  {{Huang}}, \citenamefont {{Zhao}}, \citenamefont {{Long}}, \citenamefont
  {{Wang}}, \citenamefont {{Chen}}, \citenamefont {{Yang}}, \citenamefont
  {{Liang}}, \citenamefont {{Xue}}, \citenamefont {{Weng}}, \citenamefont
  {{Fang}}, \citenamefont {{Dai}},\ and\ \citenamefont {{Chen}}}]{TaAs5}%
  \BibitemOpen
  \bibfield  {author} {\bibinfo {author} {\bibfnamefont {X.}~\bibnamefont
  {{Huang}}}, \bibinfo {author} {\bibfnamefont {L.}~\bibnamefont {{Zhao}}},
  \bibinfo {author} {\bibfnamefont {Y.}~\bibnamefont {{Long}}}, \bibinfo
  {author} {\bibfnamefont {P.}~\bibnamefont {{Wang}}}, \bibinfo {author}
  {\bibfnamefont {D.}~\bibnamefont {{Chen}}}, \bibinfo {author} {\bibfnamefont
  {Z.}~\bibnamefont {{Yang}}}, \bibinfo {author} {\bibfnamefont
  {H.}~\bibnamefont {{Liang}}}, \bibinfo {author} {\bibfnamefont
  {M.}~\bibnamefont {{Xue}}}, \bibinfo {author} {\bibfnamefont
  {H.}~\bibnamefont {{Weng}}}, \bibinfo {author} {\bibfnamefont
  {Z.}~\bibnamefont {{Fang}}}, \bibinfo {author} {\bibfnamefont
  {X.}~\bibnamefont {{Dai}}}, \ and\ \bibinfo {author} {\bibfnamefont
  {G.}~\bibnamefont {{Chen}}},\ }\bibfield  {title} {\enquote {\bibinfo {title}
  {{Observation of the chiral anomaly induced negative magneto-resistance in 3D
  Weyl semi-metal TaAs}},}\ }\href@noop {} {\bibfield  {journal} {\bibinfo
  {journal} {ArXiv e-prints}\ } (\bibinfo {year} {2015})},\ \Eprint
  {http://arxiv.org/abs/1503.01304} {arXiv:1503.01304 [cond-mat.mtrl-sci]}
  \BibitemShut {NoStop}%
\bibitem [{\citenamefont {Neupane}\ \emph {et~al.}(2014)\citenamefont {Neupane}
  \emph {et~al.}}]{WeylExp1}%
  \BibitemOpen
  \bibfield  {author} {\bibinfo {author} {\bibfnamefont {M.}~\bibnamefont
  {Neupane}} \emph {et~al.},\ }\bibfield  {title} {\enquote {\bibinfo {title}
  {Observation of a three-dimensional topological dirac semimetal phase in
  high-mobility Cd$_3$As$_2$},}\ }\href {\doibase 10.1038/ncomms4786} {\bibfield
  {journal} {\bibinfo  {journal} {Nature Communications}\ }\textbf {\bibinfo
  {volume} {5}} (\bibinfo {year} {2014}),\ 10.1038/ncomms4786}\BibitemShut
  {NoStop}%
\bibitem [{\citenamefont {Liu}\ \emph {et~al.}(2014{\natexlab{a}})\citenamefont
  {Liu} \emph {et~al.}}]{WeylExp2}%
  \BibitemOpen
  \bibfield  {author} {\bibinfo {author} {\bibfnamefont {Z.~K.}\ \bibnamefont
  {Liu}} \emph {et~al.},\ }\bibfield  {title} {\enquote {\bibinfo {title} {A
  stable three-dimensional topological dirac semimetal Cd$_3$As$_2$},}\ }\href
  {\doibase 10.1038/nmat3990} {\bibfield  {journal} {\bibinfo  {journal}
  {Nature Materials}\ }\textbf {\bibinfo {volume} {13}},\ \bibinfo {pages}
  {677--681} (\bibinfo {year} {2014}{\natexlab{a}})}\BibitemShut {NoStop}%
\bibitem [{\citenamefont {Liu}\ \emph {et~al.}(2014{\natexlab{b}})\citenamefont
  {Liu} \emph {et~al.}}]{WeylExp3}%
  \BibitemOpen
  \bibfield  {author} {\bibinfo {author} {\bibfnamefont {Z.~K.}\ \bibnamefont
  {Liu}} \emph {et~al.},\ }\bibfield  {title} {\enquote {\bibinfo {title}
  {Discovery of a three-dimensional topological dirac semimetal, Na$_3$Bi},}\
  }\href {\doibase 10.1126/science.1245085} {\bibfield  {journal} {\bibinfo
  {journal} {Science}\ }\textbf {\bibinfo {volume} {343}},\ \bibinfo {pages}
  {864--867} (\bibinfo {year} {2014}{\natexlab{b}})}\BibitemShut {NoStop}%
\bibitem [{\citenamefont {Jeon}\ \emph {et~al.}(2014)\citenamefont {Jeon} \emph
  {et~al.}}]{WeylExp4}%
  \BibitemOpen
  \bibfield  {author} {\bibinfo {author} {\bibfnamefont {S.}~\bibnamefont
  {Jeon}} \emph {et~al.},\ }\href {\doibase 10.1038/nmat4023} {\bibfield
  {journal} {\bibinfo  {journal} {Nature Materials}\ }\textbf {\bibinfo
  {volume} {13}},\ \bibinfo {pages} {851--856} (\bibinfo {year}
  {2014})}\BibitemShut {NoStop}%
\bibitem [{\citenamefont {Yi}\ \emph {et~al.}(2014)\citenamefont {Yi} \emph
  {et~al.}}]{WeylExp5}%
  \BibitemOpen
  \bibfield  {author} {\bibinfo {author} {\bibfnamefont {H.}~\bibnamefont {Yi}}
  \emph {et~al.},\ }\bibfield  {title} {\enquote {\bibinfo {title} {Evidence of
  topological surface state in three-dimensional dirac semimetal Cd$_3$As$_2$},}\
  }\href {\doibase 10.1038/srep06106} {\bibfield  {journal} {\bibinfo
  {journal} {Scientific Reports}\ }\textbf {\bibinfo {volume} {4}} (\bibinfo
  {year} {2014}),\ 10.1038/srep06106}\BibitemShut {NoStop}%
\bibitem [{\citenamefont {Azbel'}\ and\ \citenamefont
  {Kaner}(1956{\natexlab{a}})}]{AzbelKaner1}%
  \BibitemOpen
  \bibfield  {author} {\bibinfo {author} {\bibfnamefont {M.Ya.}\ \bibnamefont
  {Azbel'}}\ and\ \bibinfo {author} {\bibfnamefont {E.A.}\ \bibnamefont
  {Kaner}},\ }\bibfield  {title} {\enquote {\bibinfo {title} {The theory of
  cyclotron resonance in metals},}\ }\href@noop {} {\bibfield  {journal}
  {\bibinfo  {journal} {Zh. Eksp. Teor. Fiz.}\ }\textbf {\bibinfo {volume}
  {30}},\ \bibinfo {pages} {811} (\bibinfo {year} {1956}{\natexlab{a}})},\
  \bibinfo {note} {[Sov. Phys. JETP 3, 772 (1956)]}\BibitemShut {NoStop}%
\bibitem [{\citenamefont {Azbel'}\ and\ \citenamefont
  {Kaner}(1956{\natexlab{b}})}]{AzbelKaner2}%
  \BibitemOpen
  \bibfield  {author} {\bibinfo {author} {\bibfnamefont {M.Ya.}\ \bibnamefont
  {Azbel'}}\ and\ \bibinfo {author} {\bibfnamefont {E.A.}\ \bibnamefont
  {Kaner}},\ }\bibfield  {title} {\enquote {\bibinfo {title} {The theory of
  cyclotron resonance in metals},}\ }\href@noop {} {\bibfield  {journal}
  {\bibinfo  {journal} {Zh. Eksp. Teor. Fiz.}\ }\textbf {\bibinfo {volume}
  {32}},\ \bibinfo {pages} {896} (\bibinfo {year} {1956}{\natexlab{b}})},\
  \bibinfo {note} {[Sov. Phys. JETP 5, 730 (1956)]}\BibitemShut {NoStop}%
\bibitem [{\citenamefont {Kaner}(1958)}]{AzbelKaner3}%
  \BibitemOpen
  \bibfield  {author} {\bibinfo {author} {\bibfnamefont {E.A.}\ \bibnamefont
  {Kaner}},\ }\href@noop {} {\bibfield  {journal} {\bibinfo  {journal} {Dokl.
  Akad. Nuak. SSSR}\ }\textbf {\bibinfo {volume} {119}},\ \bibinfo {pages}
  {471} (\bibinfo {year} {1958})},\ \bibinfo {note} {[Sov. Phys. DOKL. 3, 314
  (1958)]}\BibitemShut {NoStop}%
\bibitem [{\citenamefont {Azbel'}(1960)}]{AzbelKaner4}%
  \BibitemOpen
  \bibfield  {author} {\bibinfo {author} {\bibfnamefont {M.Ya.}\ \bibnamefont
  {Azbel'}},\ }\bibfield  {title} {\enquote {\bibinfo {title} {A new resonance
  effect in metals at high frequencies},}\ }\href@noop {} {\bibfield  {journal}
  {\bibinfo  {journal} {Zh. Eksp. Teor. Fiz.}\ }\textbf {\bibinfo {volume}
  {39}},\ \bibinfo {pages} {400} (\bibinfo {year} {1960})},\ \bibinfo {note}
  {[Sov. Phys. JETP 12, 283 (1962)]}\BibitemShut {NoStop}%
\bibitem [{\citenamefont {Gantmakher}(1962)}]{AzbelKaner5}%
  \BibitemOpen
  \bibfield  {author} {\bibinfo {author} {\bibfnamefont {V.~F.}\ \bibnamefont
  {Gantmakher}},\ }\bibfield  {title} {\enquote {\bibinfo {title} {Dimensional
  effect in a metal in multiples of a certain magnetic field},}\ }\href@noop {}
  {\bibfield  {journal} {\bibinfo  {journal} {Zh. Eksp. Teor. Fiz.}\ }\textbf
  {\bibinfo {volume} {43}},\ \bibinfo {pages} {345} (\bibinfo {year} {1962})},\
  \bibinfo {note} {[Sov. Phys. JETP 16, 247 (1963)]}\BibitemShut {NoStop}%
\bibitem [{\citenamefont {Lifshitz}\ and\ \citenamefont
  {Kosevich}(1955)}]{Lifshitz}%
  \BibitemOpen
  \bibfield  {author} {\bibinfo {author} {\bibfnamefont {L.~M.}\ \bibnamefont
  {Lifshitz}}\ and\ \bibinfo {author} {\bibfnamefont {A.~M.}\ \bibnamefont
  {Kosevich}},\ }\href@noop {} {\bibfield  {journal} {\bibinfo  {journal} {Zh.
  Eksp. Teor. Fiz.}\ }\textbf {\bibinfo {volume} {29}},\ \bibinfo {pages} {730}
  (\bibinfo {year} {1955})},\ \bibinfo {note} {[Sov. Phys. JETP 2, 636
  (1956)]}\BibitemShut {NoStop}%
\end{thebibliography}
\end{document}